\documentclass[preprint,12pt]{aastex}

\shorttitle{N2D+ Depletion}
\shortauthors{Tobin et al.}
\newcommand{\mum}{\mbox{$\mu$m}}
\newcommand{\nthp}{\mbox{N$_2$H$^+$}}
\newcommand{\htdp}{\mbox{H$_2$D$^+$}}
\newcommand{\ntdp}{\mbox{N$_2$D$^+$}}
\newcommand{\nht}{\mbox{NH$_3$}}

\newcommand{\kms}{\mbox{km s$^{-1}$}}
\newcommand{\htp}{\mbox{H$_3^+$}}

\begin{document}

\title{Resolved Depletion Zones and Spatial Differentiation of \nthp\ and \ntdp\footnotemark}
\author{John J. Tobin\altaffilmark{2,3}, Edwin A. Bergin\altaffilmark{3}, Lee Hartmann\altaffilmark{3}, Jeong-Eun Lee\altaffilmark{4},
S\'ebastien Maret\altaffilmark{5}, Phillip C. Myers\altaffilmark{6}, Leslie W. Looney\altaffilmark{7}, Hsin-Fang Chiang\altaffilmark{7,8},
Rachel Friesen\altaffilmark{9}}

\begin{abstract}

We present a study on the spatial distribution of \ntdp\ and \nthp\ in thirteen 
protostellar systems. Eight of
thirteen objects observed with the IRAM 30m telescope show relative offsets
 between the peak \ntdp\ ($J=2\rightarrow1$) and \nthp\ ($J=1\rightarrow0$) emission.
We highlight the case of L1157 using interferometric
observations from the Submillimeter Array
and Plateau de Bure Interferometer of the \ntdp\ ($J=3\rightarrow2$) and \nthp\ ($J=1\rightarrow0$) 
transitions respectively.
Depletion of \ntdp\ in L1157 is clearly observed inside a radius of $\sim$2000 AU (7\arcsec) and
the \nthp\ emission is resolved into two peaks at radii of $\sim$1000 AU (3.5\arcsec),
inside the depletion region of \ntdp. Chemical models predict
a depletion zone in \ntdp\ and \ntdp\ due to destruction of \htdp\ at T $\sim$ 20 K and the evaporation
of CO off dust grains at the same temperature.
However, the abundance offsets of 1000 AU between the two species are not reproduced by chemical
models, including a model that follows the infall of the protostellar envelope.
The average abundance ratios of \ntdp\ to \nthp\ have been shown to decrease as 
protostars evolve by Emprechtinger et al., but this is the first time depletion zones of \ntdp\ have been 
spatially resolved.
We suggest that the difference in depletion zone radii for \nthp\ and \ntdp\ is caused by either
the CO evaporation temperature being above 20 K or an H$_2$ ortho-to-para ratio gradient in the inner envelope.


\end{abstract}
\footnotetext[1]{Based on observations carried out with the IRAM 30m Telescope, the IRAM Plateau de Bure Interferometer, and the Submillimeter Array.
 IRAM is supported by INSU/CNRS (France), MPG (Germany) and IGN (Spain).}
\altaffiltext{2}{Hubble Fellow, National Radio Astronomy Observatory, Charlottesville, VA 22903; jtobin@nrao.edu}
\altaffiltext{3}{Department of Astronomy, University of Michigan, Ann Arbor, MI 48109}
\altaffiltext{4}{Department of Astronomy and Space Science, Kyung Hee University, Yongin-si, Gyeonggi-do 446-701, Korea}
\altaffiltext{5}{UJF-Grenoble 1 / CNRS-INSU, Institut de Plan\'etologie et d'Astrophysique de Grenoble (IPAG) UMR 5274, Grenoble, F-38041, France}
\altaffiltext{6}{Harvard-Smithsonian Center for Astrophysics, 60 Garden Street, Cambridge, MA 02138}
\altaffiltext{7}{Department of Astronomy, University of Illinois, Urbana, IL 61801 }
\altaffiltext{8}{Institute for Astronomy, University of Hawaii at Manoa, Hilo, HI 96720}
\altaffiltext{9}{National Radio Astronomy Observatory, Charlottesville, VA 22903}
\section{Introduction}

The dense (n $\sim$10$^5$ cm$^{-3}$), cold (T$\sim$10 K) gas of starless 
or prestellar cores present one of the prime conditions
for the formation of deuterated molecules in the interstellar medium. 
These conditions are needed to allow the formation of \htdp; reactions with 
this molecule and more highly deuterated isotopomers are the primary routes for molecules to gain
molecular [D]/[H] ratios greater than the cosmic ratio of 10$^{-5}$ within T $\sim$ 10 K gas \citep{watson1974,langer1985}.
Furthermore, at these
high densities and low temperatures, CO is frozen-out onto dust grains, increasing the 
rate of deuterium fractionation in the gas-phase, as CO is a primary destroyer of \htp\ and \htdp\ \citep[e.g.][]{bergin2007}.
A key molecular tracer of dense gas where CO has frozen out is \nthp\ \citep[e.g.][]{caselli2002}. Its deuterated
counterpart \ntdp\ is found to form with abundances relative to \nthp\ of 5 - 10\% in 
starless cores, with some extreme examples having 20 - 40\% \citep{crapsi2005}.

Shortly after the gravitational collapse of a starless core begins, a Class 0 protostar
is formed within the dense core \citep{andre1993}.
This phase may last a few times 10$^5$ yr \citep{evans2009} and is accompanied by
significant luminosity evolution and likely episodic accretion \citep{young2005, kenyon1990,dunham2010}.
The combination of protostellar and accretion luminosity has a significant effect on the 
chemical abundances in the surrounding envelope \citep{lee2004}. Regions near the protostar are heated 
above 20 K (R $\la$ 1000 AU for low-mass/low-luminosity systems) and CO is evaporated off the dust grains
back into the gas phase. Then as the protostellar luminosity
increases with time, the region of CO evaporation increases as does the region where \htdp\ is converted back
into HD and \htp. Thus, cold deuterium chemistry is brought to
 a halt in the inner envelope \citep{robertsmillar2000, caselli2008,robertsmillar2007}. 

\citet{emp2009} showed that this process of a growing CO evaporation and \htdp\ destruction region will
affect the \ntdp\ to \nthp\ ratio and demonstrated how it could be used as an evolutionary indicator.
The youngest Class 0 protostars are expected to have higher \ntdp\ to \nthp\ ratios
than more evolved Class 0 protostars.
This is because the inside-out growth of the CO evaporation and \htdp\ destruction region
causes the overall abundance of \ntdp\ to drop relative to \nthp, due to the formation of
 \ntdp\ (and other deuterated species) being most efficient at the highest densities.
Thus, as the luminosity of the protostar increases with 
time, the \ntdp\ abundance will fall faster than the \nthp\ abundance. The depletion zone of \ntdp\ may
then be slightly larger ($\sim$200 AU) than that of \nthp, shown by the chemical models in \citet{emp2009}.

Depletion zones of \nthp\ have been spatially resolved previously, the most
dramatic being IRAM 04191, with a 1600 AU radius \nthp\ depletion zone \citep{belloche2004}.
Other apparent depletion zones were pointed in \citet{chen2007} and \citet{tobin2011};
depletion zones of \nht\ were also seen in some sources with \nthp\ depletion 
regions \citep{tobin2011}, as well as IRAM 04191 (Mangum et al. unpublished). 
However, most studies of deuterated molecules, in particular \ntdp\ \citep[e.g.][]{robertsmillar2007},
are primarily based on single-dish, single-point observations and do not spatially resolve or
map the emission. \ntdp\ and \nthp\ emission maps have been observed toward many starless cores and one protostellar
core \citep{crapsi2004, crapsi2005}. These studies find close agreement between \nthp\ 
and \ntdp\ distributions; however, a larger sample of \ntdp\ maps toward protostellar
sources are lacking.

In conjunction with the \nthp\ kinematics survey of \citet{tobin2011}, we simultaneously
observed \ntdp\ toward a set of thirteen low-mass protostellar systems. Furthermore, we execute 
more detailed analysis of the protostellar system L1157 where we resolve symmetric depletion zones
in both molecules using high-resolution interferometric
observations. The symmetry and likely filamentary nature of the dense protostellar envelope around
L1157 \citep[e.g.][]{gueth2003,looney2007,tobin2010a,chiang2010} make it an ideal system for examining 
the detailed spatial distribution of these species. The observations and data analysis 
are described in Section 2, the observational results of the spatial distribution
of \ntdp\ relative to \nthp\ in L1157 and twelve other systems are described in Section 3,
we compare a chemical model to L1157 in Section 4, the results are discussed and possible causes of the
spatial offset of \ntdp\ relative to \nthp\ are given in Section 5, and our 
main conclusions are given in Section 6.

\section{Observations}

This paper includes observations taken with the Submillimeter Array, the Plateau de Bure Interferometer and 
the IRAM 30m telescope. The properties of the selected sample are given in Table 1, with our sample being a subset 
of those in \citet{tobin2011}. We detail the observations and data reduction in the subsequent sections and list the relevant
details of the observations in Tables 2 and 3.

\subsection{SMA data}
L1157 was observed with the Submillimeter Array (SMA) at Mauna Kea on 2009 September 15 in sub-compact
configuration. The weather conditions were optimal with stable phases and less than 3 mm of precipitable water vapor, yielding
a zenith opacity at 225 GHz of 0.1; system temperatures were between 120 and 150 K throughout the track.
The receivers were tuned to the \ntdp\ ($J=3\rightarrow2$) transition ($\nu$=231.321635 GHz; \citet{lovas1992}) in the upper-sideband and
800 MHz of each side band was allocated to continuum; see Table 2 for a summary of observation details.
The data were edited and calibrated using the MIRIAD software package \citep{sault1995}; minimal flagging of 
uncalibrateable data was necessary (i.e. phase/amplitude jumps) and the T$_{sys}$ correction was applied, putting the
data on a Jansky amplitude scale.
The nearby quasars 1849+670 and 1927+739 were used for phase and amplitude calibration, Mars was
used for absolute flux calibration, and the quasar 3C454.3 was the bandpass calibrator. Uncertainty in absolute flux calibration is 
expected to be $\sim$10\%.

\subsection{IRAM 30m data}
Observations were taken with the IRAM 30m radio telescope on Pico Veleta using the EMIR receiver
 over several nights during 2009 October (Table 3).
For L1157, we observed the \ntdp\ ($J=2\rightarrow1$) ($\nu$=154.217096 GHz) and 
($J=3\rightarrow2$) ($\nu$=231.321966 GHz) transitions along with 
the \nthp\ ($J=3\rightarrow2$) ($\nu$=279.511701 GHz) and ($J=1\rightarrow0$) ($\nu$=93.173404 GHz) 
transitions, see Table 3 for details. We also observed
11 other sources in the \nthp\ ($J=1\rightarrow0$) and \ntdp\ ($J=2\rightarrow1$) transitions 
simultaneously. System temperatures were between 100 K and 140 K at 93 GHz and between 
110 K to 230 K at 154 GHz, see Table 3.

All data were taken in frequency-switched On-The-Fly (OTF) mapping mode. 
For L1157 in particular, the \ntdp\ ($J=3\rightarrow2$) and
\nthp\ ($J=3\rightarrow2$) lines were mapped in a 90\arcsec\ $\times$ 90\arcsec\ 
box centered on the protostar and \nthp\ ($J=1\rightarrow0$)
and \ntdp\ ($J=2\rightarrow1$) were observed in a 4\arcmin\ $\times$ 3\arcmin\ box centered 30\arcsec\ south of
the protostar. Data for the additional sources were generally observed in  3\arcmin\ $\times$ 3\arcmin\ regions
centered near the protostar. The OTF maps were constructed by fully scanning each source
in both the north-south and east-west directions (5\arcsec\ steps between scan legs); the orthogonal mapping
directions minimize striping
in the final map. The maps were repeated when necessary for a higher signal-to-noise ratio.
We conducted system temperature measurements (calibration scans) about every 10
minutes and about 2 hours were needed to complete each map. 
The pointing offset was measured once every two hours, preceeding the observation of each source.  The offsets were generally $\sim$5\arcsec\
and this offset was stable throughout an observing session, having variations of only $\sim$2\arcsec\ from one
pointing scan to the next. The RMS pointing accuracy of the telescope is $\sim$2\arcsec\ and our observed 
variations are consistent with this. The receiver offsets between
the 3 mm and 2 mm bands were $\sim$2\arcsec\ and had been measured only a month before the
data were taken; such a small offset will not affect our results.

The initial calibration of the OTF data to the antenna temperature scale and CLASS
data format was performed automatically on-site by the Multichannel Imaging and calibration software
for Receiver Arrays (MIRA)\footnote[2]{http://www.iram.fr/IRAMFR/GILDAS} package. The default
amplitude calibration is expected to be accurate within $\sim$10\%. Further data reduction was done using CLASS
(part of GILDAS\footnotemark[2]). For all molecular lines observed, the frequency switched spectra were folded
and baseline subtracted using a second order polynomial.
We then reconstructed the spectral map on a grid such that the FWHM of the beam
was spanned by 3 pixels and each pixel is the average of all measurements within
the FWHM of the beam, see Tables 2 and 3, as well as \citet{tobin2011} for additional details. Lastly,
the emission maps were put on the T$_{mb}$ scale where T$_{mb}$ = F$_{eff}$T$_{A}^*$/B$_{eff}$. F$_{eff}$ is
the forward efficiency, accounting for losses by the secondary support structure, and B$_{eff}$ is the main beam
efficiency; values for F$_{eff}$ and B$_{eff}$ are dependent on frequency and given in Table 2.

\subsection{PdBI data}

L1157 was observed with the Plateau de Bure Interferometer in the \nthp\ ($J=1\rightarrow0$) 
transition in the D and C configurations in 2009 June and 2009 November. The data were edited, calibrated, and mapped
using the CLIC and MAPPING components of the GILDAS software package. The absolute flux was derived from observations of
MWC349, assuming a flux density of 1.15~Jy at 3~mm for that source; uncertainty in the absolute
flux is $\sim$10\%. More details on the data reduction are given in \citet{tobin2011}.

\subsection{Single-Dish and Interferometric Combination}
As an additional step, we combined the single-dish \nthp\ data from the 30m with the PdBI data.
We regridded the single-dish data to the phase center of the PdBI observations and resampled
the velocity axis to correspond to the 0.125 \kms\ velocity resolution of the PdBI
data. We did this using the 
MAPPING component of GILDAS, performing a joint deconvolution of the single-dish and interferometric data;
GILDAS provides built-in routines for combining the 30m data. The resulting 
image provides substantially better extended structure sensitivity than
the default interferometric image shown in \citet{tobin2011}.

We also explored the possibility of combining the single-dish \ntdp\ data with the SMA \ntdp\ data for L1157. 
Unfortunately, the single-dish \ntdp\ ($J=3\rightarrow2$) data had a lower signal-to-noise ratio than the
SMA data, prohibiting a useful 
combination of the data.

\subsection{Data Analysis and Hyperfine Fitting}
The nuclear spin of nitrogen causes both \nthp\ and \ntdp\ to have a hyperfine emission line
spectrum.
The multiple hyperfine components and line ratios of \nthp\ \citep{caselli1995,keto2010}
 and \ntdp\ \citep{dore2004} enable us to simultaneously determine
the line-center velocity, linewidth, optical depth, and excitation temperature at
each point in the map.

CLASS was used to fit the hyperfine emission line structure, which can fit multiple lines
given the optically thin line ratios and velocity offsets. The lines were fit at each pixel in the spectral
map using a semi-automated routine. This routine used the integrated intensity map (zeroth moment)
of the line emission, calculated from the total line intensity in a velocity range and multiplied
by the channel velocity width. The pixels with at least a 3$\sigma$ detection of \nthp\ or \ntdp\
were selected and a CLASS script to fit the hyperfine line structure at each pixel was generated;
the resultant fit parameters (line-center velocity, linewidth, 
optical depth, and excitation temperature) were written out to a text table.
This criteria does not require 3$\sigma$ detections of all line components and there were cases
in which only the strongest hyperfine lines are detected. Bad fits resulting from inadequate signal-to-noise
were removed to generate the final table of pixels with good fits.
The typical failure mode of the fitting is excessively large
linewidths ($>$1.5 \kms) and/or optical depths of 30, making the discrepant points easy to identify. 

The integrated intensity, excitation temperature, and optical depths from line fitting are 
then used to calculate the column densities of the data following \citet{goldsmith1999}. 
The column density maps generally trace regions where the integrated intensity was at least $\sim$6$\sigma$ and
usually 10$\sigma$. We use the excitation temperature of the \nthp\ ($J=1\rightarrow0$) 
in the calculation of \ntdp\ column density when the \ntdp\ ($J=2\rightarrow1$) optical depth is less than 1.0.
This is because the excitation temperature cannot be well-constrained if the lines are optically
thin. If the excitation temperature of \ntdp\ ($J=2\rightarrow1$) is actually lower than that of \nthp\ ($J=1\rightarrow0$),
the \ntdp\ abundances would systematically underestimated. When the \ntdp\ excitation temperatures could
be constrained from the hyperfine fitting, they are comparable but generally lower than the values fit for \nthp\ ($J=1\rightarrow0$).
The uncertainties in optical depth and excitation temperature lead to column density 
uncertainties of order $\sim$ 20\% to 30\%, in addition to 
a possible 10\% amplitude calibration uncertainty. A few case have $>$100\% uncertainty in the column densities due to 
large uncertainties in the optical depth.

We also computed spectra averaged within a 12\arcsec\ radius 
(roughly the beamwidth at of the IRAM 30m at 93 GHz) centered on the position
of each protostar. We used these spectra to derive average abundances and column
densities toward the protostars for comparison with 
previous single-pointing studies \citep[e.g.][]{robertsmillar2007,emp2009}. 

\section{Observational Results}

We obtained simultaneous \ntdp\ and \nthp\ mapping of twelve protostellar envelopes, as well as interferometric
data in both species for L1157. The best example of spatially resolved \ntdp\ and \nthp\ depletion is found toward
L1157 in the interferometric data. We will discuss this source in detail first before presenting the rest of the sample.

\subsection{L1157}

The SMA continuum and \ntdp\ ($J=3\rightarrow2$) data (4.3\arcsec$\times$7.1\arcsec\ resolution)
 for L1157 are shown in Figure 1, overlaid on the \textit{Spitzer}
IRAC 8 \micron\ image. The 1.3 mm continuum data show that the source is strongly dominated by
the point source emission coincident with the protostar, arising from the dense inner protostellar envelope.
There is also extension along the outflow direction, likely due to warm gas associated with 
the inner outflow cavity. The outflow is nearly within the plane of the sky \citep{gueth1997} having
an inclination of $\sim$85\degr. At lower intensity levels, the continuum emission is found to be extended along the
dense envelope that was detected in extinction at 8 \mum\ \citep{looney2007,tobin2010a}
and in \nthp/\nht\ molecular line emission \citep{chiang2010,tobin2011}. This extension
along the extinction lane was previously seen in the continuum with 
single-dish bolometer mapping at 850 \micron\ and 
1.3 mm \citep{gueth2003}, but not in interferometric continuum data 
\citep{gueth1997, gueth2003, chiang2010,chiang2012}. The 3 mm 
continuum image from the PdBI is centrally peaked with no indication of 
extension along the envelope (Figure 1), consistent with \citet{chiang2010}.
 The detection of emission extended along the envelope with the SMA likely owes to the high-sensitivity of the 
observation, the short uv-spacings in subcompact configuration, and the brighter
dust continuum emission at 1.3 mm as compared to 3 mm.

The \ntdp\ ($J=3\rightarrow2$) integrated intensity map (Figure 1), unlike the 
continuum data, is double-peaked, roughly symmetric about 
the protostar, with a local minimum coincident with the position of the protostar. 
The peaks are at radii of $\sim$ 2000 AU (7\arcsec) from the protostar and 
the emission closely correlates with opaque regions of
the 8 \micron\ extinction feature. In contrast, the \nthp\ integrated intensity 
map from the PdBI+30m has emission peaks that are located at R$\sim$1000 AU (3.3\arcsec) from the protostar 
(Figure 1); the \ntdp\ emission peaks are 1000 AU further out. The double-peaked \nthp\ emission
appears at the same location in both the PdBI+30m image and the PdBI-only image from \citep{tobin2011}.
The symmetry about the protostar in \ntdp\ and \nthp\ is striking, a 
strong indication that the protostar is affecting the abundance distribution of these molecules.

We also conducted single-dish mapping 
of \nthp\ in the ($J=3\rightarrow2$) and ($J=1\rightarrow0$) transitions and
\ntdp\ in the ($J=3\rightarrow2$) and ($J=2\rightarrow1$) transitions. 
The integrated intensity maps of these lines are shown in Figure \ref{L1157-30m} and
spectra of all four lines toward the location of the protostar 
are shown in Figure \ref{L1157-30m-spectra}. The \nthp\ 
maps both appear centrally peaked and the ($J=3\rightarrow2$) 
transition appears slightly elongated along the envelope axis, consistent
with the resolved PdBI data. Both \ntdp\ maps show a double-peaked feature, 
coincident with the widely separated emission peaks observed in the 
SMA data.

 The single-dish data having centrally peaked \nthp\ ($J=3\rightarrow2$) emission,
double-peaked \ntdp\ ($J=3\rightarrow2$) and ($J=2\rightarrow1$) rule-out any excitation anomaly, 
optical depth, or interferometric filtering being responsible
for the apparent central depletion of \ntdp\ emission around L1157. If the higher temperatures
at smaller radii were causing the apparent central deficit of \ntdp\ ($J=3\rightarrow2$), then 
the \nthp\ ($J=3\rightarrow2$) should have appeared to be similarly double-peaked. Optical depth is ruled out
because fitting the hyperfine transitions of \ntdp\ finds them to be optically thin; the optically
thin nature of the emission is apparent in the spectra shown in Figure 3 where the satellite lines emission
lines are quite faint relative to the main lines. Therefore, we conclude that the deficit
of \ntdp\ emission toward the protostar and the offset between interferometric \nthp\ and
\ntdp\ peaks reflects a reduction of \ntdp\ abundance relative to \nthp\ in the inner envelope.

We used the interferometric \nthp\ and \ntdp\ maps to construct column density maps of each
species in order to construct a ratio map of \ntdp\ to \nthp. The \nthp\ data have the single-dish 
data combined with the interferometer data to correct for missing flux. The \ntdp\ data have
resolved out some flux, apparent from the negative contours in Figure 1; however, the total interferometric
flux is comparable to the single-dish flux indicating that most emission is recovered in the SMA map. 
The SMA data were sampled with the same pixel scale and phase center as the PdBI data and
we calculated the column densities for each transition following \citet{goldsmith1999}. The excitation temperature
for \ntdp\ was taken from the hyperfine fits of \nthp\ ($J=1\rightarrow0$) at each point because the 
\ntdp\ ($J=3\rightarrow2$) was optically thin and the excitation temperature could not be determined from hyperfine fitting.
Figure \ref{L1157-ratiomap} shows the SMA \ntdp\ ($J=3\rightarrow2$) emission contours overlaid on the
PdBI+30m \nthp\ ($J=1\rightarrow0$) map and shows the column density ratio of \ntdp\ relative to \nthp\ in the
adjacent panel. We then take one dimensional radial cuts of the column densities and ratio and plot
them in Figure \ref{L1157-abund}. 

The line parameters for the average \nthp\ and \ntdp\ spectra
 within a 12\arcsec\ radius around the protostar are given in Table 4 and the 
column densities and ratios are given in Table 5. We calculate that 
L1157 has 3 times the level of \ntdp\ as measured by \citet{emp2009}, but a similar \nthp\ column density. This may result from us having mapped
the area surrounding the protostar, while \citet{emp2009} appear to have only taken spectra in single pointings centered on the protostar.
In which case, the 16\arcsec\ beam at the frequency of \ntdp\ ($J=2\rightarrow1$) would not have sampled the region of peak 
\ntdp\ emission with only one pointing. The single-dish ratio is also an order of magnitude higher than the minimum observed in
the interferometer data, mainly due to the \nthp\ column density
being about and order of magnitude larger in the interferometer data. The lower column density in the single-dish data can 
be attributed to beam dilution and if we consider the column densities and ratios at 3600 AU radii (12\arcsec), then 
the interferometer data are consistent with the single-dish.

\subsection{Other Sources}
While we have given significant attention to L1157, observations of a larger sample show that 
\ntdp\ and \nthp\ are often spatially differentiated and L1157 is not a special case.
Figure 6 shows the \nthp\ emission maps of the full sample overlaid on the IRAC 8 \mum\ images,
compared with the \ntdp\ emission maps. Figure 7 shows the column density maps
of \nthp, \ntdp, and the ratio of \ntdp\ to \nthp\ for HH211 as an example and only the ratio
maps are shown for the rest of the sample in Figure 8. The column densities are calculated by the method 
described in Section 2, using the integrated intensity corrected for optical depth
and excitation temperature. The parameters of the \nthp\ and \ntdp\ line
fits are given in Table 4 and our column densities and ratios are given in Table 5. Our calculated
column densities and ratios are consistent with those in \citet{emp2009} within about a factor of two. The
differences likely result from our exact centering on the protostar and our 12\arcsec\
radius for averaging the spectra to derive an overall column density and abundance ratio.
The sources are briefly discussed individually below.

\textbf{IRAS 03282+3035} -- This source has a very dense envelope extended in the north-south direction,
with nearly coincident \nthp\ and \ntdp\ emission peaks. The
ratio map indicates that the peak \ntdp\ abundance is offset from the protostar and \nthp\ peak on one side only
and to the south by 8\arcsec. This offset seems real and is larger than the 2\arcsec\ receiver offset.

\textbf{HH211} -- This protostar was the most heavily deuterated object in 
\citet{emp2009} and this is true of our sample as well (see Table 5), with a comparable
ratio of \ntdp\ to \nthp. However, those measurements were
toward the protostar position; the peak emission of \nthp\ and \ntdp\ is offset to the southwest. Despite the \nthp\ emission
peak being offset from the protostar, the column density peak is centered on the protostar (Figure 7). 
The \ntdp\ column density peak is, however, offset southwest
and the peak level of \ntdp/\nthp is 0.28 at (-9\arcsec, 9\arcsec).

\textbf{L1521F} -- This protostar was classified as a Very Low 
Luminosity Object (VeLLO) by \citet{bourke2006} and was previously 
regarded as a starless core \citep{crapsi2004}. The \ntdp\ emission 
peak may be slightly offset to the north relative to the \nthp;
the overall spatial distribution of \ntdp\ appears quite flat; 
these maps are consistent with those of \citet{crapsi2004}.
The column density ratio maps indicates that the \ntdp\ abundance 
may be reduced toward the central protostar.

\textbf{IRAS 04325+2402} -- This protostar is a multiple Class I source in Taurus, 
but with a large, dense core in close proximity, approximately located at (10, 25) in Figure 6
and offset from the protostar by $\sim$63\arcsec\ \citep{scholz2010,hoger2000}.
The peak \nthp\ emission is found to be north of the protostar, coincident with the nearby 
dense core seen in 8 \micron\ extinction, with emission extending southward toward the 
protostars. The \ntdp\ emission is all located north of the protostar, coincident 
with the \nthp\ peaks. The detection of \ntdp\ toward the protostar listed in Table 4 is
marginal.

\textbf{L1527} -- L1527 is also in Taurus and \citet{emp2009} did 
not detect \ntdp\ emission toward this source. However, we find 
that the \ntdp\ emission is concentrated north of the protostar 
by $\sim$70\arcsec\ (10000 AU) and north of the \nthp\ peak by 
$\sim$35\arcsec\ (5000 AU), see Figure 6. Despite the large offset of \ntdp\ from the protostar, the submillimeter
emission and 8 \micron\ extinction are also more extended on the north-side
of the protostar and coincide with the \ntdp\ emission \citep{tobin2010a, chandler2000}. 
\citet{robertsmillar2007} also detected 
\ntdp\ toward this source in a lower resolution study with comparable 
column density. Despite the Class 0 status of L1527, it is very depleted in 
\ntdp\ relative to others in the sample. This could result from the
wide outflow cavities increasing the overall envelope temperature and driving
down the deuteration.

\textbf{RNO43} -- This source was the most distant in the sample, associated with Orion. Nonetheless,
we did find that the \ntdp\ emission was spatially offset from the \nthp\ emission, with the \ntdp\ mostly 
on the west side of the protostar. The column density and ratio map is quite noisy due to the faintness of \ntdp; moreover,
the abundance of \ntdp\ relative to \nthp\ may be overestimated due to beam dilution of the \nthp\ map, given that RNO43 has
the lowest \nthp\ column density in the sample.

\textbf{HH108 IRS} -- In the case of this protostar, the \ntdp\ and \nthp\ emission are both centered to the southeast of the protostar.
Unfortunately, the signal-to-noise of the \ntdp\ was too poor to make a useful column density ratio map.

\textbf{HH108 MMS} -- This is the nearby neighbor of HH108 IRS, but is apparently less 
evolved and lower luminosity. Its ratio
of \ntdp\ to \nthp\ is about 4 times greater than HH108 IRS, consistent with the
apparent youth of this protostar relative to HH108 IRS. The \ntdp\ emission from his 
protostar might be offset, but the noisy map makes this indefinite.

\textbf{L483} -- L483 is the strongest \nthp\ source in the sample and most optically thick. The \ntdp\ emission peak
is offset by $\sim$20\arcsec\ west of the protostar, with the highest \ntdp\ to \nthp\ ratio at the location of peak \ntdp\ emission.
However, the column density ratio map (Figure 8) indicates that there could be higher column density \ntdp\  
coincident with the highest column density of \nthp.

\textbf{L673} -- The subregion of L673 mapped coincides with L673-SMM2 as identified by \citet{visser2002}, harboring
a small clustering of young stars. The sources marked in 
Figure 6 and 8 correspond to the Class 0/I sources 13, 27, and 28 in \citet{tsitali2010}. The \ntdp\ emission 
is clearly not peaked at the same location as the \nthp\ emission and the column density is substantially reduced.
The average \nthp\ to \ntdp\ ratios are fairly constant and low for the three identified sources.
However, the northern most source (28) may be at the location of higher \ntdp\ abundance as shown in the column density and ratio maps, but
this is uncertain.

\textbf{L1152} -- L1152 is comprised of two cores seen in \ntdp\ 
and \nthp\ emission linked by an apparent filament of material. The protostellar
core in the south has strong \ntdp\ and \nthp\ emission, but 
offset from the protostar. The peak \ntdp\ to \nthp\ ratios in the south (protostellar) and
north (starless) are 0.09 and 0.11 respectively.

\textbf{L1157} -- We performed the same analysis on L1157 as 
the rest of the sample for completeness, the drop in \ntdp\ abundance
is not apparent in the single-dish ratio map (Figure 8), but there is a
slight depression of emission in the \ntdp\ ($J=2\rightarrow1$) map. 
The average \ntdp\ to \nthp\ ratio of 0.12 is consistent with 
averaging the minimum and maximum abundance ratio from the interferometer data.

\textbf{L1165} -- Similar to L483, the \ntdp\ emission is also 
shifted along the outflow in L1165. The emission does seem to extend back
toward the \nthp\ peak that is southeast of the protostar. 
The small area of \ntdp\ emission makes a ratio map uninteresting, but the source
has one of the lowest ratios of \ntdp\ to \nthp\ from the single-point analysis; 
however this could be affected by beam dilution.

Including L1157, eight of the thirteen sources show spatial shifts in the
peak emission or column density of \ntdp\ relative to \nthp.
Although L673 is included, it could be argued that it is different since its environment is more complex due to having several young stars
within a 0.1 pc diameter region. It is also important to note that we are limited by resolution in many cases, so some sources
could have shifts that are not detectable in the 30m data. The \nthp\ and \ntdp\ shifts can be regarded with 
high confidence because the data were taken simultaneously and the receivers are aligned to
within 2\arcsec\ accuracy. The relative offset between \nthp\ and \ntdp\ peaks  
are tabulated in Table 5. In addition to \ntdp\ being shifted relative to \nthp, there are also offsets
between the peak emission of \nthp\ and the protostars in six of the thirteen sources 
(HH108 IRS, L1152, L1165, HH211, L1527, and IRAS 04325+2402).

While spatial differentiation between the species is quite prevalent,
the location of the peaks relative to the protostar
is often not as well-ordered as L1157. We observe a trend in the sample
in which the apparently youngest sources, as classified from their cold bolometric temperatures ($T_{bol}$) (Table 1)
(HH211, HH108MMS, IRAS 03282+3035, and L1521F) do not show as strong spatial 
differentiation between \ntdp\ and \nthp. The more evolved
protostars (with warmer $T_{bol}$ and wide outflows \citep{arce2006})  have the largest spatial 
offset between \ntdp\ and \nthp\ (L1527 and IRAS 04325+2402).

We also compare the \ntdp\ to \nthp\ ratios to the bolometric luminosity and temperature in Figure \ref{lboltbol}. 
There is not a clear trend in bolometric luminosity, only that more luminous sources tend to have less deuteration while lower luminosity
sources may be more highly deuterated or not. There does appear to be a trend of decreasing \ntdp\ to \nthp\ ratio with
increasing bolometric temperature, indicating that the deuteration may be decreasing with protostellar evolution. However, there is significant
scatter at low bolometric temperatures.

\subsection{Kinematics of \ntdp\ versus \nthp}

The difference in relative spatial distributions between \ntdp\ and \nthp\ in
most systems is an indication that these molecules are not tracing the same material 
throughout the envelope. This suggests that the kinematic
structure probed by these tracers is likely different. We compare the kinematics of the two tracers 
for L1157 in Figures \ref{L1157-sd-kinematics} \& \ref{L1157-int-kinematics}, using both
single-dish and interferometric kinematics respectively. 
For L1157, we notice a similar large-scale velocity gradient in both \nthp\ and \ntdp; 
however, near the protostar, where the \ntdp\ abundance drops, there are clear differences. 
The line-of-sight velocities measured from \ntdp\ on the west side of the envelope appear systematically larger 
by about 0.1 \kms\ as compared to the \nthp\ data; this is present in both 
the single-dish and interferometric data. This region is located near the blueshifted
outflow cavity and could be due to \nthp\ being entrained by the outflow \citep{tobin2011},
skewing the line profile toward blue-shifted velocities. Since the \ntdp\ abundance is peaked
at larger radii, it may not be affected by the outflow as strongly as \nthp. Thus, the \ntdp\ could
be tracing an increased velocity gradient due to rotation and/or infall \citep{tobin2011,tobin2012}
that is masked by the outflow interaction with \nthp.

The linewidth of \nthp\ is also found to be larger than \ntdp\ within 10\arcsec\ 
of the protostar in both the single-dish and interferometer data. Thus,
the signature of the outflow-envelope interaction is not as pronounced in \ntdp\ due to its lower abundance on
scales less than 2000 AU. This is a clear indication that on small-scales, the \ntdp\ is not
tracing the same gas as \nthp. Moreover, this demonstrates that \nthp\
 is able to probe closer to the protostar and disk forming region of the envelope 
during this stage of evolution. 

\section{Chemical Model Comparison}

L1157 represents an ideal case to test chemical models for a protostellar source due to the
higher resolution data, excellent signal-to-noise, and well-known orientation in the plane of
the sky. Moreover, the relative symmetry of the filamentary
envelope \citep{tobin2010a}, \nthp, and \ntdp\ emission on scales less than 5000 AU 
 make chemical modeling of this source more tractable.
We will first review the critical chemical processes at work, affecting the formation and abundance
of \nthp\ and \ntdp.

\subsection{Relevant Chemical Processes}

Deuterium chemistry in the cold (T$\sim$ 10 K) interstellar medium begins with the formation
of H$_2$D$^+$, dominated by the reaction
\begin{equation}
{\rm H_3^+}  +  {\rm HD}  \rightleftharpoons  {\rm H_2D^+} + {\rm H_2} + \Delta {\rm E}
\end{equation}
where $\Delta E$ $\sim$ 230 K \citep{herbst1982}.
At low temperatures (T $\la$ 20 K) the forward reaction will dominate causing H$_2$D$^+$ to have an abundance approaching
that of H$_3^+$ in cold, dense cores \citep{langer1985}. The high abundance of reactive
H$_2$D$^+$ enables subsequent deuterated molecules further down the reaction chain 
to have abundances elevated above the cosmic [D/H] $\sim$ 10$^{-5}$, 
including D$_2$H$^+$ and D$_3^+$ \citep{caselli2008, pagani2009}.
The abundance of H$_2$D$^+$ is limited by recombination with electrons and reactions with other molecules.
 The back reaction can also become non-negligible at T $\sim$ 20 K, leading to decreased deuteration even
before CO has been liberated from the dust \citep{emp2009}. 

 At T $<$ 20 K, most CO is
frozen-out onto dust grains at the densities of protostellar cores 
(n$\sim$10$^5$ cm$^{-3}$), but once the protostar raises the temperature 
to $\ga$ 20 K, the CO ice sublimates back to the gas phase. 
When CO is in the gas phase, all the available
carbon not locked into dust grains will react, effectively 
shutting off the deuterium chemistry as initiated
by \htdp; an additional route to deuterium fractionation is
 available through C\htdp at T $>$ 30 K, but this does not cause subsequent 
\ntdp\ formation. Furthermore, the H$_2$ ortho-to-para ratio
of the cloud can affect the overall deuterium fractionation; 
ortho-H$_2$ is in the J=1 state with an energy above the ground of 170 K. This has enough energy
to lower the potential barrier of the back reaction 
of Equation 1 even at low temperatures, making the overall level of deuteration sensitive 
to the H$_2$ ortho-to-para ratio \citep{flower2006, pagani2009}.

The primary formation route for \nthp\ is the gas phase reaction
\begin{equation}
{\rm N_{2}} + {\rm H_3^{+}} \rightarrow {\rm N_{2}H^{+}}
\end{equation}
and \ntdp\ is understood to form via the same process, but with H$_2$D$^+$ as the reactant. Thus, the ability
to form \ntdp\ is constrained by the conditions in which H$_2$D$^+$ exists. Both \ntdp\
and \nthp\ are destroyed by reactions with CO, therefore, they are only expected to be present 
with significant abundance in regions
where CO has depleted. In low-mass/luminosity protostellar systems, the temperatures only rise above 20 K at radii
less than 1000 AU. Thus, the results from chemical modeling lead us to 
expect \ntdp\ and \nthp\ to trace the same gas where CO is frozen-out onto dust grains \citep[e.g.][]{emp2009, pagani2009}.

\subsection{Chemical Model}

We compare our data with the evolutionary chemical model of \citet{lee2004}. 
Since L1157 has formed a protostar, a model that accounts for the infalling 
material as well as the evolution of radiative heating is important.
This model endeavors
to simulate the chemistry of an isolated protostellar cloud through the prestellar phase as approximated
by a Bonnor-Ebert sphere and then transition to the protostellar phase using the inside-out collapse model \citep{shu1977}
with an initially isothermal temperature of 10 K. The protostellar phase includes
luminosity evolution following the prescription of \citep{young2005} with radiative transfer calculated by DUSTY, but
not including episodic accretion bursts. The chemistry of the infalling 
gas parcels is calculated discretely and the inward radial motion is tracked within the model.

Since the chemical model calculates a 1D radial abundance profile, we took several steps to make a more
realistic observational comparison. We used the input radial density profile from the chemical model
and the calculated abundance profile to construct a circular envelope midplane.
We then summed the density along the line of sight through the envelope to calculate the
predicted column density distribution in order to compare with the observations.
Since L1157 is likely a filament rather than a sheet \citep{tobin2010a}, we also tapered the circular sheet
by a Gaussian along the assumed line of sight with scale height parameters of 500, 750, 1000, and 2000 AU in order
to simulate a filamentary density distribution. However, we note that the chemical abundances calculated for a
1D density profile will not exactly map to a filamentary envelope, but should be a reasonable approximation.
Our goal was not to fit L1157 in detail with the model, but to look for a general agreement with the 
observed radial distribution and abundance ratio of the two species.

 In Figure \ref{L1157-abund-model}, we show a plot of the \ntdp\ to \nthp\ column density ratios 
as a function of radius from the model. We clearly see the drop in abundance of the two
species at radii less than 1000 AU due to reactions with CO and their abundances peak at intermediate radii
between 1000 and 5000 AU, with decreasing abundances at larger radii. The model that best
fits the abundance ratio of L1157 has evolved for 
300,000 yr after collapse and the luminosity of the model (5 L$_{\sun}$) is comparable to what is observed.
While the abundance ratio of \ntdp\ relative to \nthp\
matches the data reasonably well, the column density and abundance peaks of \nthp\ and \ntdp\ are nearly coincident with each other, 
inconsistent with the data shown in Figure \ref{L1157-abund-model}.
Earlier and later timesteps of the model also show coincident peaks between these two molecules.
Thus, predicted ratio of \ntdp\ and \nthp\ is consistent with the model, but the absolute abundances/column densities
are inconsistent. 

It is important to point out that the column density peak of \ntdp\ is in agreement with the observed
\ntdp\ column density peak and the \nthp\ column density peak of the model is what disagrees with the observations.
The inability of the model to reproduce the abundance peak offsets between
the \ntdp\ and \nthp\ may indicate that the models may not be capturing some physical process;
we will discuss this further in the following section.

\section{Discussion}
It has been previously shown that the depletion of \ntdp\ relative to \nthp\ is expected
 in protostars \citep{robertsmillar2007,emp2009}, but the $\sim$1000 AU
spatial differentiation between the peaks of the two species is not matched by our chemical modeling, nor
that of \citet{emp2009}.
Moreover, the only other study to recognize the spatial offset of \nthp\ and \ntdp\ was \citet{friesen2010}
with observations of the protocluster Oph B2. However, such spatial offsets have not been observed toward 
individual/isolated protostellar systems previously. Spatially resolved observations of deuterated
molecules relative to their normal counterparts offer crucial tests to chemical models and 
further our understanding of the evolution of deuterated molecules in protostellar systems.

\subsection{\nthp\ Depletion Zones}

As pointed out in the introduction, a \nthp\ depletion zone had been 
observed previously toward the Class 0 protostar IRAM 04191 \citep{belloche2004}. This
source is recognized as a Very Low Luminosity Object or VeLLO,
having an internal luminosity of $<$~0.1 L$_{\sun}$. The depletion zone in IRAM 04191
has a radius of 1600 AU, larger than that of L1157 and other protostars observed
by \citet{chen2007} and \citet{tobin2011}. However, the disappearance
of \nthp\ could be explained by either freeze-out onto dust grains, as apparently
observed in the pre-stellar core B68 \citep{bergin2002}, or the destruction of
\nthp\ due to CO being released from dust grains as a result of protostellar heating 
\citep[e.g.][]{lee2004}. Such a large depletion zone in IRAM 04191 produced
via \nthp\ destruction from evaporated CO would not work given the low system luminosity and
 \citet{belloche2004} suggested freeze-out of N$_2$ onto dust grains as the likely mechanism for
\nthp\ depletion. Alternatively, \citet{lee2007} constructed an episodic accretion model
that could also explain the size of the depletion zone with a recent burst of luminosity
in IRAM 04191. L1157 and the other protostars known to have resolved \nthp\ depletion zones are 
of order 5 to 10 L$_{\sun}$. These protostars could evaporate CO from dust grains out to 
sufficiently large radii with equilibrium heating to explain the size of the
depletion zones.

\subsection{Cause of \nthp\ and \ntdp\ Emission Offsets}

The inability of the chemical model used here \citep{lee2004} and others in the literature \citep{emp2009} 
to reproduce the relative offset of the column density peaks
for both species is troubling at first. However, this may not mean that our
understanding of the chemical processes forming and destroying the molecules is wrong.
The presence of the central protostar, raising the gas and dust temperatures above $\sim$10 K
introduces significant complexities to the chemistry within $\sim$2000 AU of the
 protostar, as compared to the pre-stellar phase. Moreover, since we are considering the
radial abundances of two molecules, we must determine whether it is the \ntdp\ or
\nthp\ that is problematic.

As outlined in the previous section, 
the reaction that forms \htdp, leading to the formation of
\ntdp, is only active at temperatures less than 20 K; at greater temperatures the 
\htdp\ reacts with H$_2$ and is converted back to \htp\ and HD (Equation 1).
Furthermore, T $\sim$ 20 K is also the temperature at which CO is 
evaporated off of dust grains and can freely react with
and destroy \ntdp\ and \nthp. A C$^{18}$O map of L1157 detects an emission peak at the protostar and very 
little CO at radii larger than 1200 AU \citep{jorgensen2007}, 
consistent with the location of the \nthp\ abundance peaks. 
CO will destroy \ntdp\ as quickly as it destroys \nthp,
so reactions with CO will affect the abundance distribution of both molecules in the same manner. 
In order to have an abundance peak of \nthp\ inside that of \ntdp, one needs to 
have a range of radii where the \htdp\ abundance is reduced and CO is not significantly
present in the gas phase. Figure \ref{n2dp-schematic} shows a schematic 
diagram of the spatial zones in L1157 where \nthp, \ntdp, and the two molecules governing
the spatial distribution of emission are located (CO and \htdp). Of 
these molecules, only observations of \htdp\ are lacking.

The relevant processes in the protostellar envelope 
that could affect the spatial distribution of molecules are protostellar heating, outflows, and infall.
\citet{emp2009} presented chemical models that considered protostellar heating and the
 \htdp\ abundances becoming reduced at radii somewhat larger
than CO sublimation; however, the shifts of \ntdp\ peak abundances were only $\sim$200 AU 
outside of the \nthp\ peak and unable to explain the 1000 AU 
offset in L1157. Replenishment of \nthp\ and \ntdp\ via
infall from the outer envelope is built into \citet{lee2004} model, but \nthp\ 
and \ntdp\ remain coincident at all time steps. \citet{tobin2011} showed that the protostellar
outflow is entraining \nthp\ at the 1000 AU scale; however, such entrainment would affect \ntdp\ in the 
same way that it affects \nthp. Moreover, \ntdp\ does not appear to be significantly affected by the 
outflow as evidenced by its narrower linewidth than \nthp\ (Figure \ref{L1157-sd-kinematics}).

Another possibility considered was episodic accretion \citep[e.g.][]{kenyon1990, dunham2010}.
If there was an accretion outburst, it could temporarily heat the envelope 
to T $\ga$ 20 K out to $\sim$2000 AU. Then all the \htdp\ could have been converted 
back to \htp\ and HD. Moreover, CO would be liberated from the dust
grains destroying the \nthp\ and \ntdp\ present \citep{lee2007,visser2012}.
After the luminosity decreases and CO freezes back out, \nthp\ would be able to form directly, 
while \ntdp\ would have to wait until \htdp\ has reformed. Thus, we considered the possibility
that \nthp\ could reform faster than \ntdp\ since it is a one-step process 
and \ntdp\ takes two steps to form. While this scenario is attractive, work by
 \citet{lee2007, visser2012} shows that the chemistry is governed by the CO 
freeze-out timescale after an accretion outburst and \ntdp\ and \nthp\ should reform in lockstep.
Thus, episodic accretion alone cannot adequately account for the observed
spatial distribution of \nthp\ and \ntdp.

An additional scenario we considered is an H$_2$ ortho-to-para (OPR) 
gradient in the envelope. The steady-state OPR
in cold (T $\sim$ 10 K) clouds is calculated to be $\sim$2.7$\times$10$^{-3}$ 
\citet{flower2006}, whereas H$_2$ forms with OPR = 3 from statistical
equilibrium. OPR values larger than steady-state would decrease the deuteration by enabling the
back reaction of \htdp\ formation to be active (Equation 1). 
Since the protostar likely went through a cold starless phase with a duration
of 10$^4$ to 10$^6$ yr the H$_2$ OPR throughout the cloud was likely
around the equilibrium value before collapse \citep{parise2011}. Once the protostar forms
and begins to heat the inner envelope, the OPR ratio could increase; \citet{sipila2010}
shows a sharp increase at T $>$ 15 K. Such an OPR ratio gradient going into the warm 
inner envelope could explain the large region of \ntdp\ depletion. However, it is unclear
if this could reproduce the observations given that the 
OPR is still quite low ($\sim$10$^{-4}$).  There is also a question of timescales
given that the steady-state timescale in \citet{sipila2010} is $\sim$10$^5$ yr, longer than
the infall timescale for material to fall in from 2000 AU to 1000 AU 
($\sim$ 10$^4$ yr assuming a 0.5 $M_{\sun}$ protostar). While there are some problems 
with this scenario, it cannot be ruled out.

A final scenario to produce the observed 
abundance offsets is a CO evaporation temperature
is higher than 20 K. The temperatures 
for \htdp\ destruction and CO evaporation are quite
close and if the CO evaporation temperature was even just a few
 degrees higher than 20 K, this could be enough to 
explain the observed emission offsets, as was true in the \citet{emp2009} models. We illustrate this in 
Figure \ref{n2dp-schematic}, where if the CO remained
mostly frozen-out between 20 to 25 K, this could explain 
the emission offsets by enabling \nthp\ to be present in this 
warmer region at smaller radii than \ntdp. It is possible that the CO evaporation temperature
could vary since the physical state of CO ice can be much more complex. 
The CO may be mixed with other ices: \nht, CO$_2$,CH$_3$OH, H$_2$O, etc.
The binding energy of CO to other ices tends to be strong 
than CO to CO \citep{collings2004,fayolle2011}, therefore ice mixtures having slightly
larger evaporation temperatures could explain the observations. 
However, it is uncertain how mixed the ice is because
some species tend to freeze-out more quickly, forming ice 
layers rather than a single mixed layer \citep{oberg2011}.
Aside from the caveat of unknown ice mixtures, a slightly 
altered evaporation temperature of CO ice due to
mixing would explain our observations quite simply without 
needing episodic accretion and/or H$_2$ OPR gradients. But,
episodic accretion could cause more ice mixing and conversion of CO ice to CO$_2$
may build an ice mixture with a large enough binding energy to remain frozen at T $>$ 20 K \citep{kim2012}.

In summary, the abundance peak offsets between \ntdp\ and \nthp\
could be explained by either an H$_2$ OPR gradient in the inner envelope
or an increased CO evaporation temperature due to ice mixtures. The chemical model
of \citet{lee2004} does not calculate a temperature dependent H$_2$ OPR, which 
may explain why it does not reproduce the abundance peak offset between \ntdp\ and \nthp. At the same time,
the model also does not allow for ice mixtures to increase the evaporation temperature CO, which could
cause the \nthp\ peak to move inward. However, the chemical model does have its \ntdp\
column density/abundance peak at roughly the same radius as the observations, this offers some
evidence for the increased CO evaporation temperature scenario, i.e. the deuterium chemistry
produced the expected results for \ntdp\ but \nthp\ is divergent.
Though, we cannot conclusively distinguish between these two scenarios as both seem likely to 
occur and could be simultaneously contributing to make the shift in abundance peaks stronger.

\subsection{Envelope and \ntdp/\ntdp\ Emission Morphology}

An additional effect to consider is the viewing angle and three dimensional 
envelope density structure in relation to the observed \ntdp\ emission morphology.
In the case of L1157, there is evidence that the envelope is filamentary rather than
sheet-like \citep{tobin2010a}. A filamentary envelope can have two effects
on the apparent \nthp\ and \ntdp\ depletion zones. First, the depletion zones
are likely more apparent in a filamentary geometry than in a axisymmetric sheet, due to less emission 
superimposed along the line of sight in than an axisymmetric object; this is demonstrated in the 
column density plots in Figure \ref{L1157-abund-model}. Also, the orientation of a filament
within the plane of the sky can change the apparent size of the depletion regions; however, viewing angle effects would only 
make the depletion regions smaller and not larger. Thus, no complicated viewing angle effects
 could give the illusion of large
\ntdp\ depletion zones outside those of \nthp. 
In the case of L1157, we cannot attribute the observed morphology to observational bias 
due to the symmetry of the depletion zones and abundance peaks.

The rest of the sample is not as well-ordered as L1157 in terms of envelope structure 
and/or emission morphology. L1527, L1165, L1152, HH108 IRS, HH211, and IRAS 04325+2402 have \nthp\
emission peaks that are not coincident with the protostar, but the \ntdp\ emission is never
peaked closer to the protostar that \nthp. L483 and L1165 show  
\ntdp\ emission peaked in the direction of the outflow rather than perpendicular to it.
In these cases of complex emission morphology, the heating of the protostar and the envelope structure
are both contributing to the complex emission morphology. L483 shows \nthp, \nht, and 850 \micron\ emission that are oriented at a 45\degr\
angle to the outflow direction \citep{fuller2000,shirley2000}, roughly the same direction as the \ntdp\ emission.
Therefore, the \ntdp\ emission apparently extended along the outflow is simply shifted in the direction
of the high-density envelope structure and the highest density material is simply not perpendicular to the outflow.

Moreover, the envelopes with less complex structure near the protostar in 8 \micron\ extinction and \nthp\ 
(i.e. L1521F, IRAS 03282+3035, HH211, and L1157) do not have irregular \ntdp\ distributions. These sources, except
for L1157, appear to have \ntdp\ emission peaks possibly shifted to one side of the envelope. 
Thus, we can conclude that if the envelope has highly irregular
structure, then we expect this to create an irregular distribution of \ntdp.
Furthermore, the frequent offsets of \ntdp\ emission from the protostars, even though it may
be coincident with the \nthp\ peak, means that \ntdp\ is not tracing the densest material in protostellar cores.
Thus, caution must be exercised when interpreting the kinematics of the deuterated species without
observations of both as deuterated species cannot be relied upon to trace the densest and/or same
material as non-deuterated in the presence of temperature gradients.

\subsection{Chemical Evolution}

It is important to stress that even though \ntdp\ and \nthp\ are not always
spatially coincident as previously assumed and indicated by chemical models, our 
results do not disagree with the generalized picture of chemical 
evolution in protostellar cores \citep[e.g.][]{lee2004, robertsmillar2007,emp2009}.
Specifically, the reduced abundance of \ntdp\ relative to \nthp\ as protostar evolve is consistent
with our data as shown in Figure \ref{lboltbol} and the previous work; moreover, the spatial offsets of \ntdp\ relative 
to \nthp\ appear to be a product of this chemical evolution and
contribute to the relative decrease of \ntdp\ relative to \nthp. 
Given the two scenarios we outline for the cause of the peak offsets, further
chemical modeling that accounts for infall, H$_2$ OPR gradients, and changes in the CO evaporation
temperature is needed to fully understand the exact cause. Sensitive, spatially-resolved
observations of both C$^{18}$O and \htdp\ could determine if the 
increased evaporation temperature is likely, as we would expect to see a gap between the two species if this
were the case.

\section{Summary and Conclusions}

We have presented a detailed study of the relative spatial distribution of \ntdp\ and \nthp\ in
Class 0/I protostellar envelopes from interferometric observations of L1157 and single-dish data of 
thirteen sources (L1157 included). We find that \nthp\ and \ntdp\ often do not have the same
column density distribution, given that the emission peaks of the two species are not spatially coincident, contrary
to predictions by previous chemical models.

1. The depletion regions of both \nthp\ ($J=1\rightarrow0$) and \ntdp\ ($J=3\rightarrow2$) in the envelope
around the Class 0 protostar L1157 are well-resolved in interferometric
observations with the PdBI and SMA respectively. Both \ntdp\ and \nthp\ are double-peaked
with the \ntdp\ peaks at R$\sim$2000 AU, outside the peaks of \nthp\ which are at R$\sim$1000 AU.
Follow-up single-dish observations from the IRAM 30m of \nthp\ ($J=3\rightarrow2$), \ntdp\ ($J=3\rightarrow2$),
and ($J=2\rightarrow1$) confirm the spatial offsets of the species.

2. Subsequent observations of additional protostellar envelopes in \nthp\ ($J=1\rightarrow0$) 
and \ntdp\ ($J=2\rightarrow1$) with the IRAM 30m show frequent offsets between the emission
peaks of \ntdp\ and \nthp\ in eight of thirteen sources (L1157 included). Some sources without
offsets could be due to a lack of spatial resolution; moreover, the additional sources with spatial
offsets are not as well-ordered as L1157.

3. The large ($\sim$1000 AU) offset in abundance peaks of \nthp\ and \ntdp\ is not matched by the
dynamic chemical model, nor previously published static models from \citep{emp2009}. In such models,
the abundance ratio of \ntdp\ and \nthp\ falls due to heating by the protostar, but
the abundance peaks remain closely coincident.

4. We suggest that the emission offsets are best explained by either an increased CO
evaporation temperature due to ice mixtures or an H$_2$ ortho-to-para ratio gradient. In
the case of an increased CO evaporation temperature, there would be a region where \htdp\
is not present in the gas phase, but CO is still frozen-out, enabling \nthp\ have its 
emission peak be inside of the \ntdp\ peak. However, an H$_2$ ortho-to-para ratio gradient 
would appear observationally similar and both effects may be happening.

We thank the referee Paola Caselli for helpful comments that improved
the final manuscript. The authors wish to acknowledge useful discussions with K. Oberg.
We thank the SMA staff for carrying out the observations, IRAM 30m staff for
allowing us to use the E330 receiver with its interim local oscillator in the Fall of 2009 and 
assistance with the observations. The Submillimeter Array is a joint project between 
the Smithsonian Astrophysical Observatory and the Academia Sinica Institute of 
Astronomy and Astrophysics and is funded by the Smithsonian Institution and the 
Academia Sinica. J.Tobin acknowledges support provided by NASA through Hubble Fellowship 
grant \#HST-HF-51300.01-A awarded by the Space Telescope Science Institute, which is 
operated by the Association of Universities for Research in Astronomy, 
Inc., for NASA, under contract NAS 5-26555. The National Radio Astronomy Observatory 
is a facility of the National Science Foundation operated under cooperative agreement 
by Associated Universities, Inc.

\begin{small}
\bibliographystyle{apj}
\bibliography{ms}
\end{small}

\begin{figure}
\begin{center}
\includegraphics[scale=0.3,angle=-90]{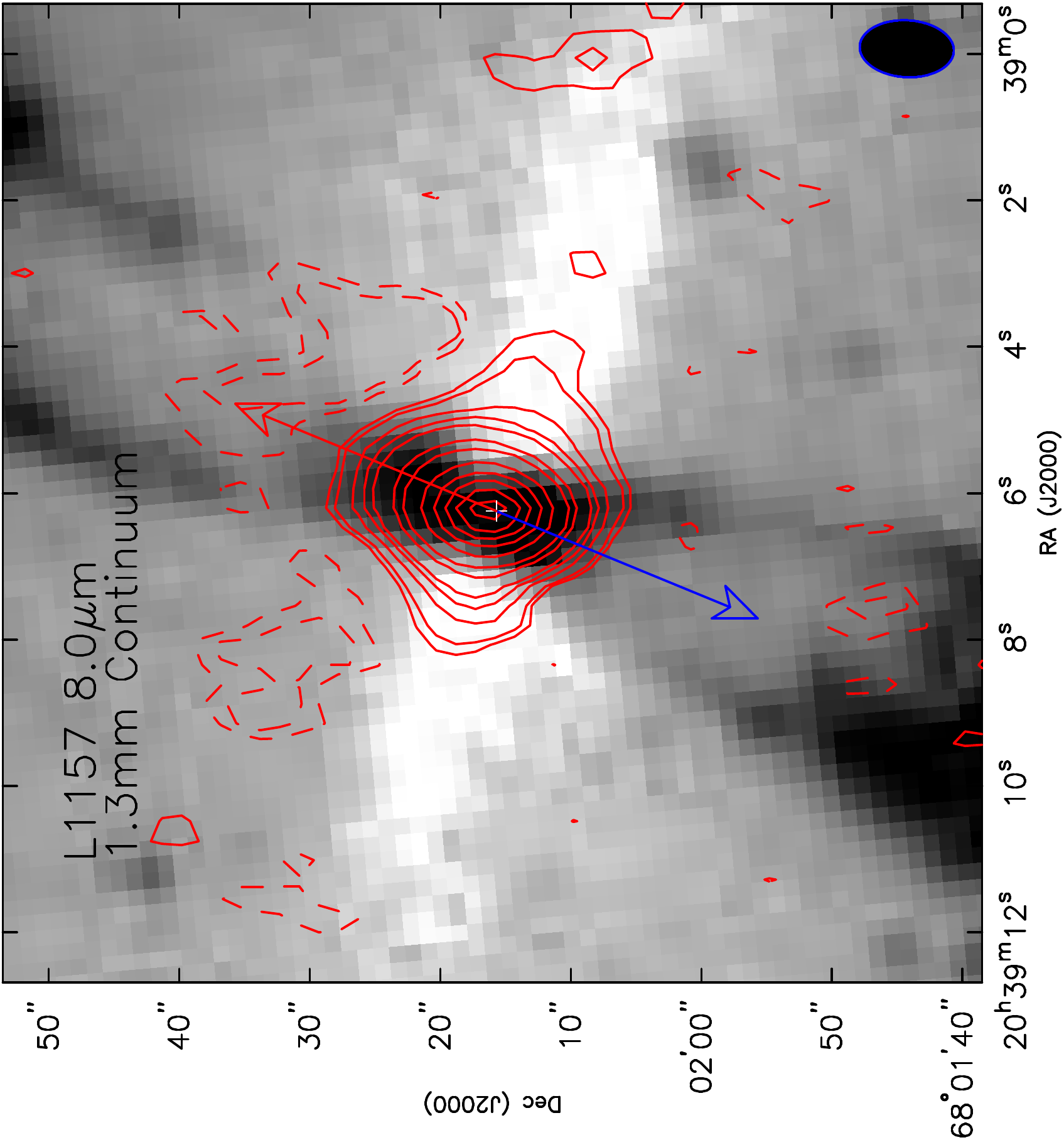}
\includegraphics[scale=0.3,angle=-90]{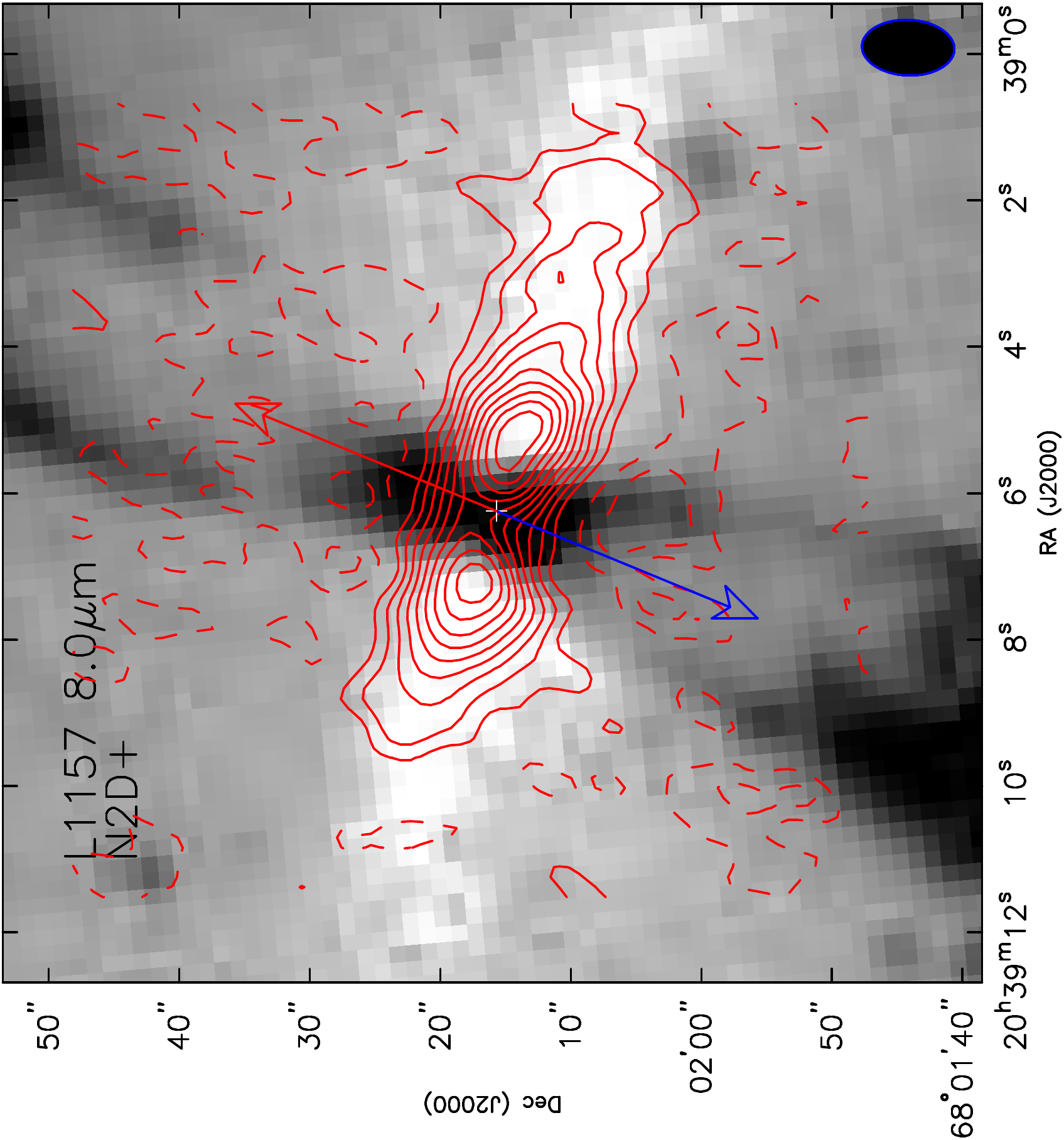}
\includegraphics[scale=0.3,angle=-90]{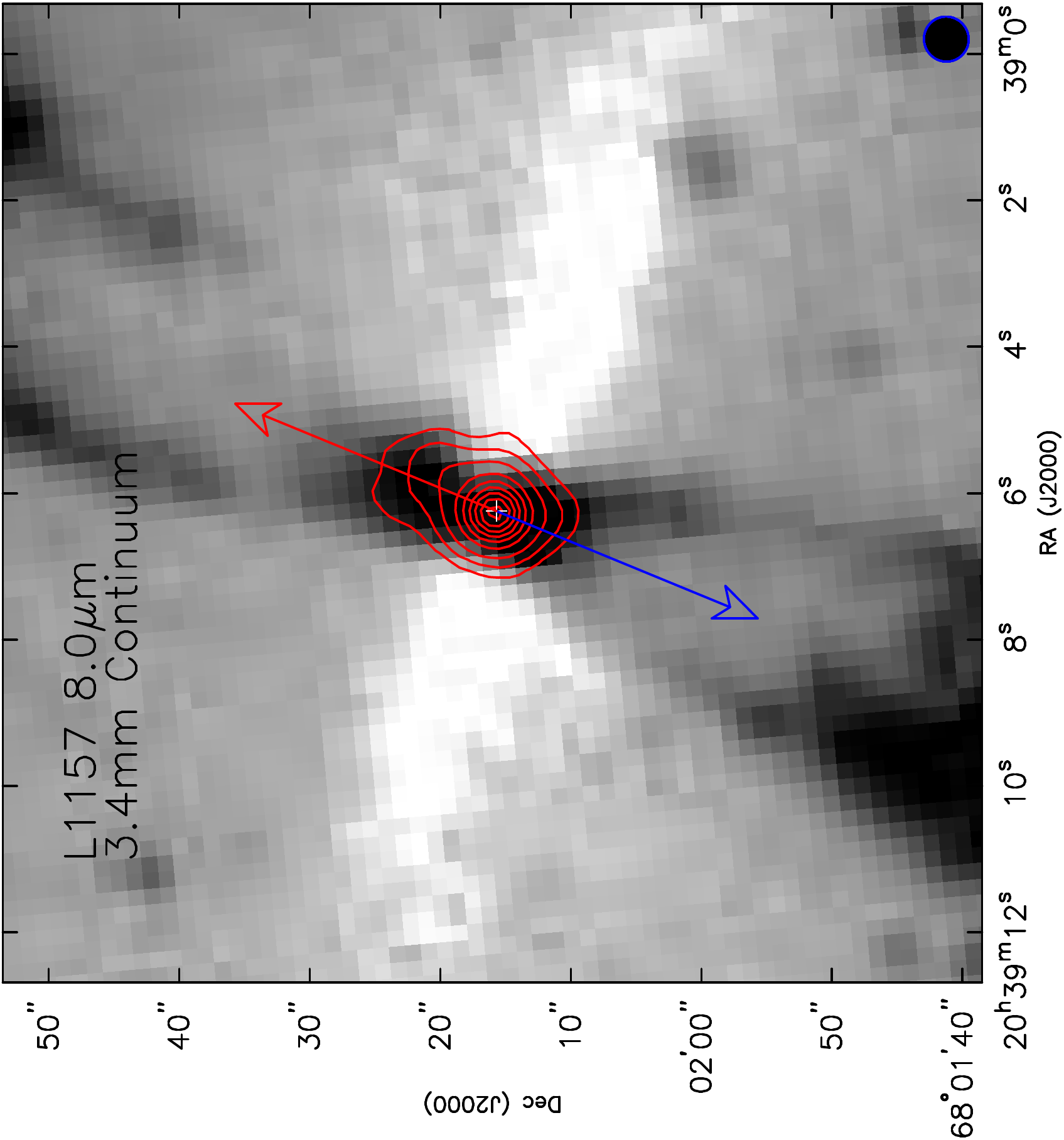}
\includegraphics[scale=0.3,angle=-90]{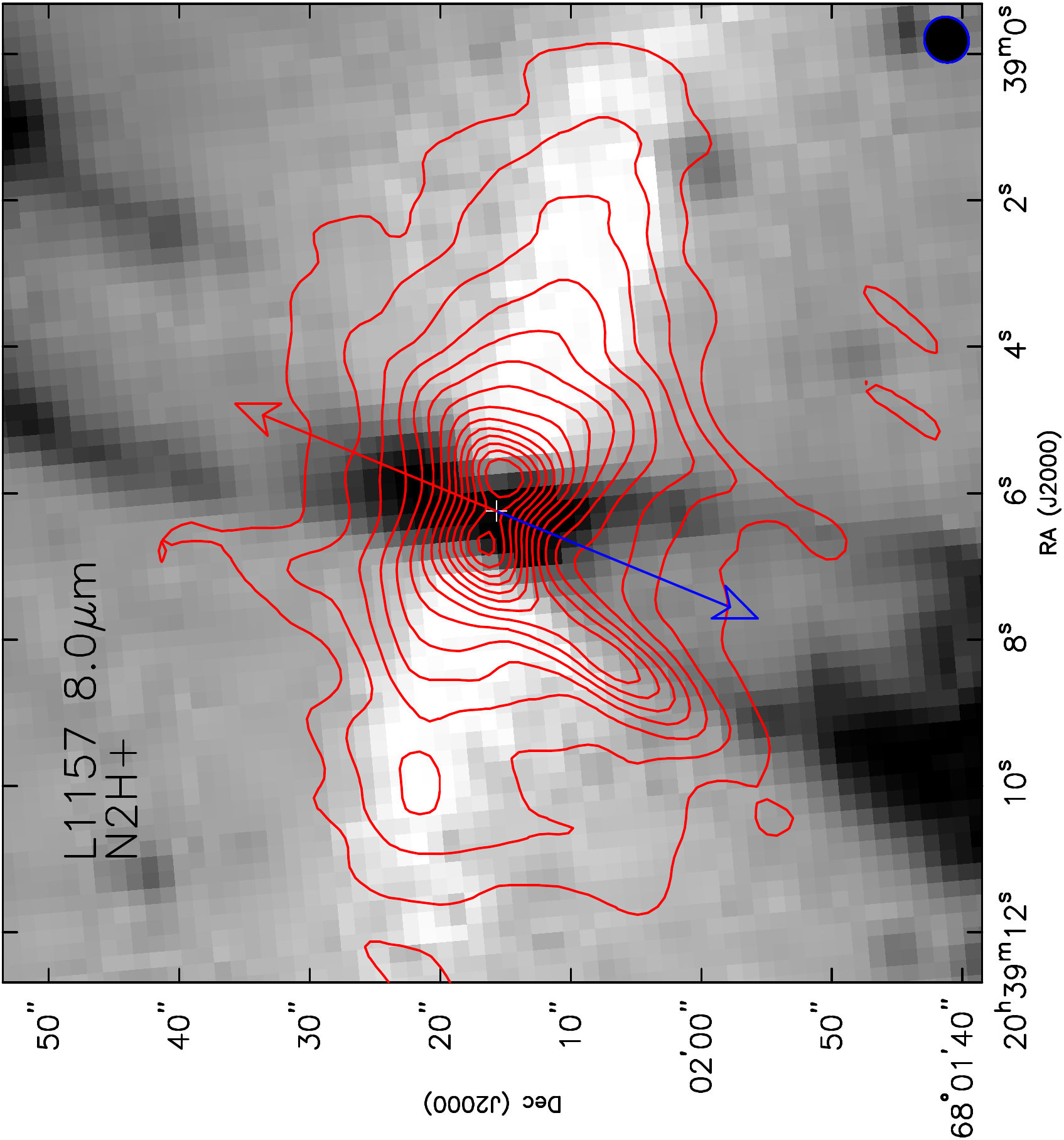}
\end{center}
\caption{L1157-- Data from the SMA and PdBI overlaid on the IRAC 8~\micron\ image (inverse grayscale).
The \textit{top left panel} shows the SMA 1.3 mm continuum emission (contours) which peaks on the protostar but
is also extended at low surface brightness levels along the envelope viewed in 8~\micron\ extinction;
contours are $\pm$2$\sigma$, $\pm$3, 6, 9, 12, 22, 30, 50, ..., 150$\sigma$, where $\sigma$ = 2.16 mJy beam$^{-1}$.
The \textit{top right panel} shows the SMA \ntdp\ ($J=3\rightarrow2$) integrated intensity emission, the contours
start at and increase in units of $\pm$3$\sigma$, where $\sigma$ = 0.07 K \kms. The \ntdp\ emission shows 
a local minimum on the protostar position and peaks at R $\sim$ 2000 AU. 
The \textit{bottom left panel} shows the PdBI 3.4 mm continuum emission (contours) which peaks on the protostar;
contours are $\pm$2$\sigma$, $\pm$3, 6, 12, 24, 36, ..., 96$\sigma$, where $\sigma$ = 0.24 mJy beam$^{-1}$.
The \textit{bottom right panel} show the PdBI + IRAM 30m \nthp\ ($J=1\rightarrow0$) integrated intensity emission;
the contours start at and increase in units of $\pm$30$\sigma$, where $\sigma$ = 0.042 K \kms. The \nthp\ emission also shows a
local minimum at the location of the protostar, but the peak emission is located at R$\sim$1000 AU and inside
the peak \ntdp\ emission. The SMA beam is 7.1\arcsec\ $\times$ 4.3\arcsec\ and the PdBI beam is 3.4\arcsec\ $\times$ 3.3\arcsec.
}
\end{figure}

\begin{figure}
\begin{center}
\includegraphics[scale=0.75,angle=-90]{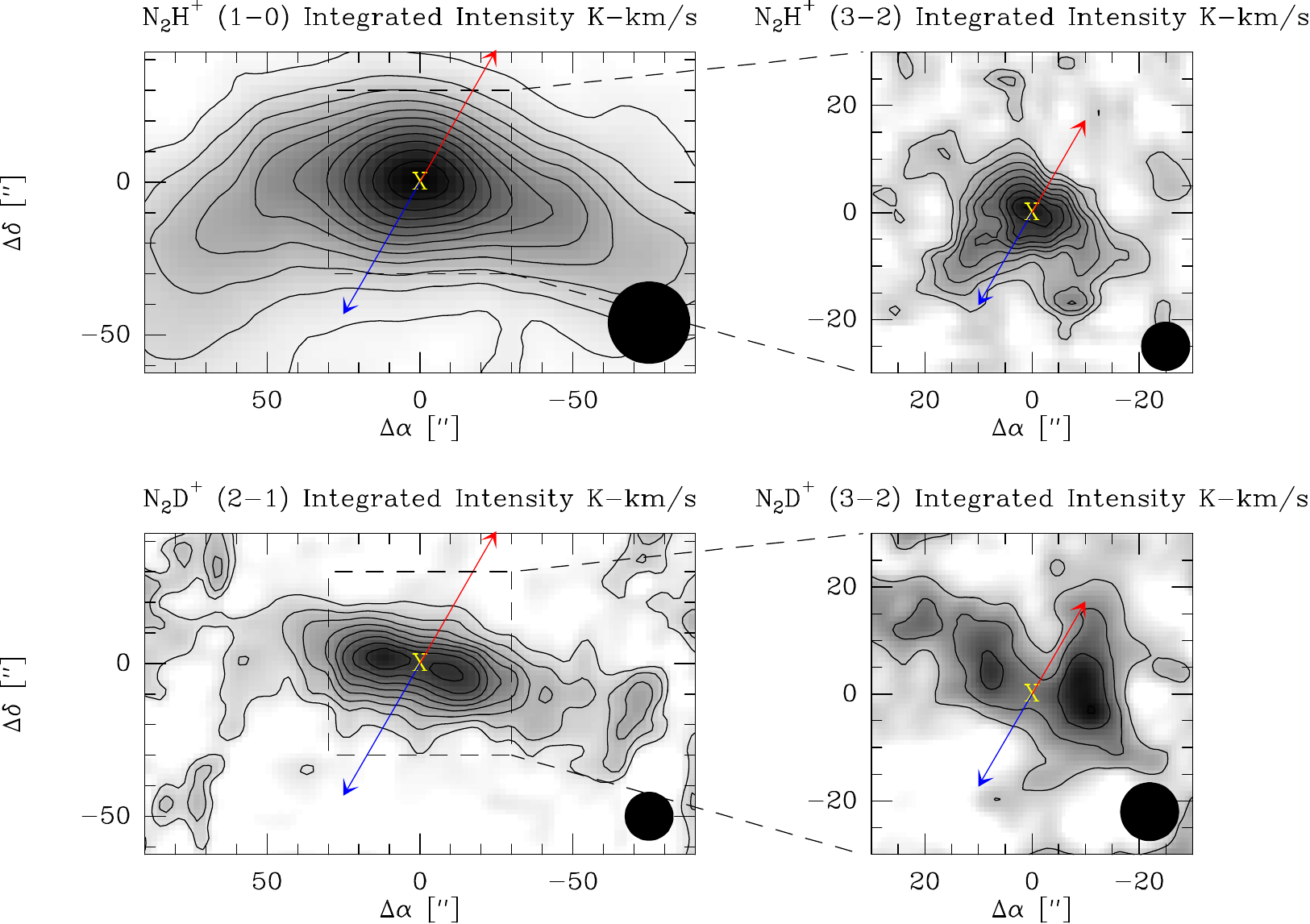}
\end{center}
\caption{L1157-- Single-dish mapping from the IRAM 30m. The \textit{top panels} show the
 \nthp\ ($J=1\rightarrow0$) and ($J=3\rightarrow2$) emission, respectively. The \textit{bottom panels}
show the \ntdp\ ($J=2\rightarrow1$) and ($J=3\rightarrow2$) emission. The \nthp\ emission is centrally peaked
and the \nthp\ ($J=3\rightarrow2$) appears elongated normal to the outflow, consistent with the double-peaked
\nthp\ ($J=1\rightarrow0$) map from the PdBI (Figure 1). The \ntdp\ emission then appears double-peaked
in both transitions, consistent with the SMA \ntdp\ observations. 
The contours and beam in each panel are drawn as follows: 
\nthp\ ($J=1\rightarrow0$)  10$\sigma$, 30$\sigma$, 50$\sigma$, ..., and $\sigma$=0.027 K \kms, 27\arcsec\ beam;
\nthp\ ($J=3\rightarrow2$)  4$\sigma$, 6$\sigma$, ..., and $\sigma$=0.14 K \kms, 9\arcsec\ beam;
\ntdp\ ($J=2\rightarrow1$) 2$\sigma$, 4$\sigma$, ..., and $\sigma$=0.05 K \kms, 16\arcsec\ beam;
\ntdp\ ($J=3\rightarrow2$) 2$\sigma$, 4$\sigma$, ..., and $\sigma$=0.12 K \kms, 11\arcsec\ beam.
}
\label{L1157-30m}
\end{figure}

\begin{figure}
\begin{center}
\includegraphics[scale=0.75]{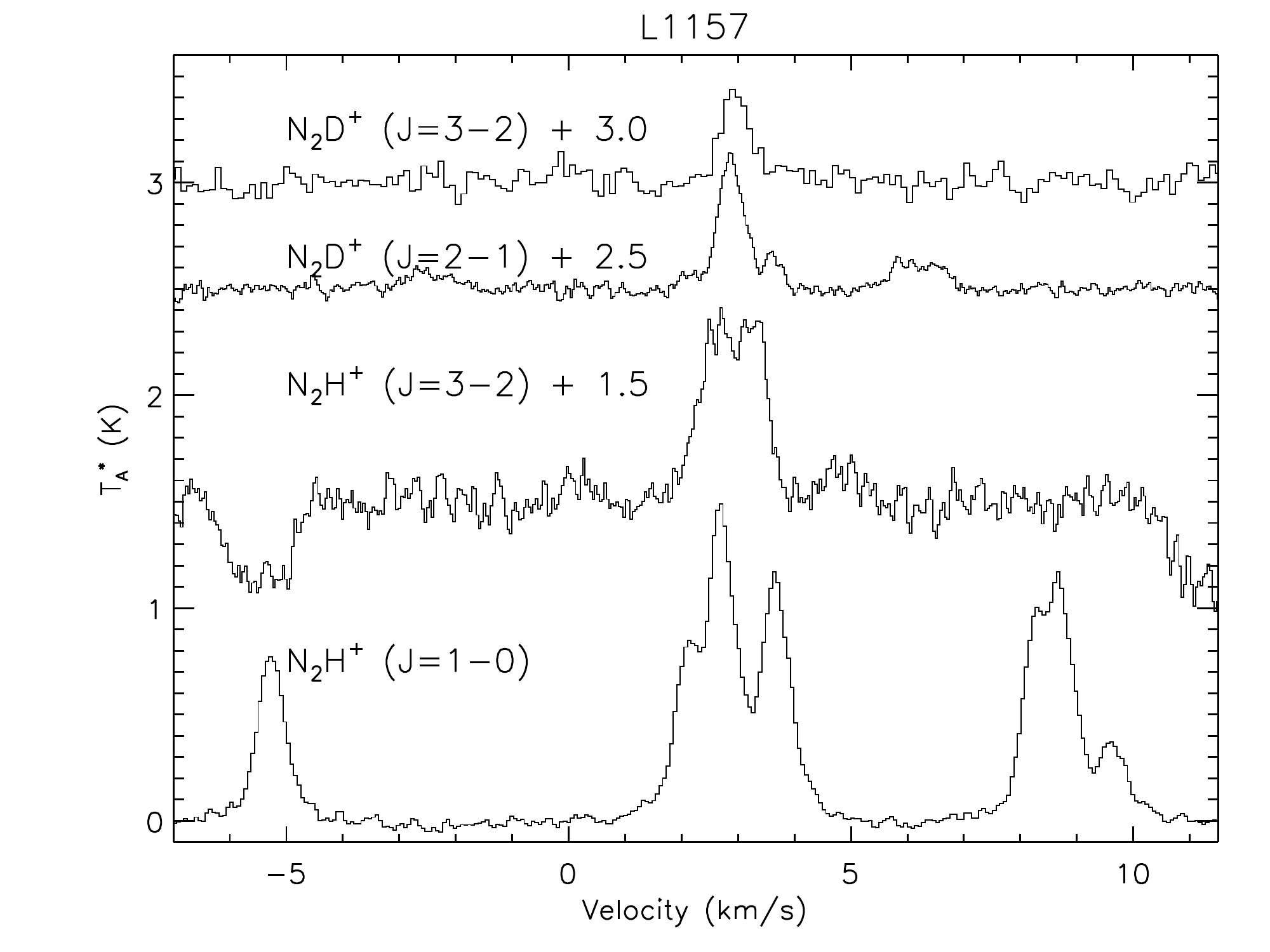}
\end{center}
\caption{Example spectra of \nthp\ and \ntdp\ taken toward L1157. The spectra are averaged over an aperture radius
of 12\arcsec\ centered on the protostar position. Faint hyperfine lines are visible in the \nthp\ ($J=3\rightarrow2$)
at 0 and 5 \kms\ and in \ntdp\ ($J=2\rightarrow1$) at -3 and 6 \kms. The signal-to-noise is too low to see
hyperfine lines in \ntdp\ ($J=3\rightarrow2$). The negative features in the \nthp\ ($J=3\rightarrow2$) at -5 and -12 \kms\
are frequency switching artifacts.
}
\label{L1157-30m-spectra}
\end{figure}

\begin{figure}
\begin{center}
\includegraphics[scale=0.8]{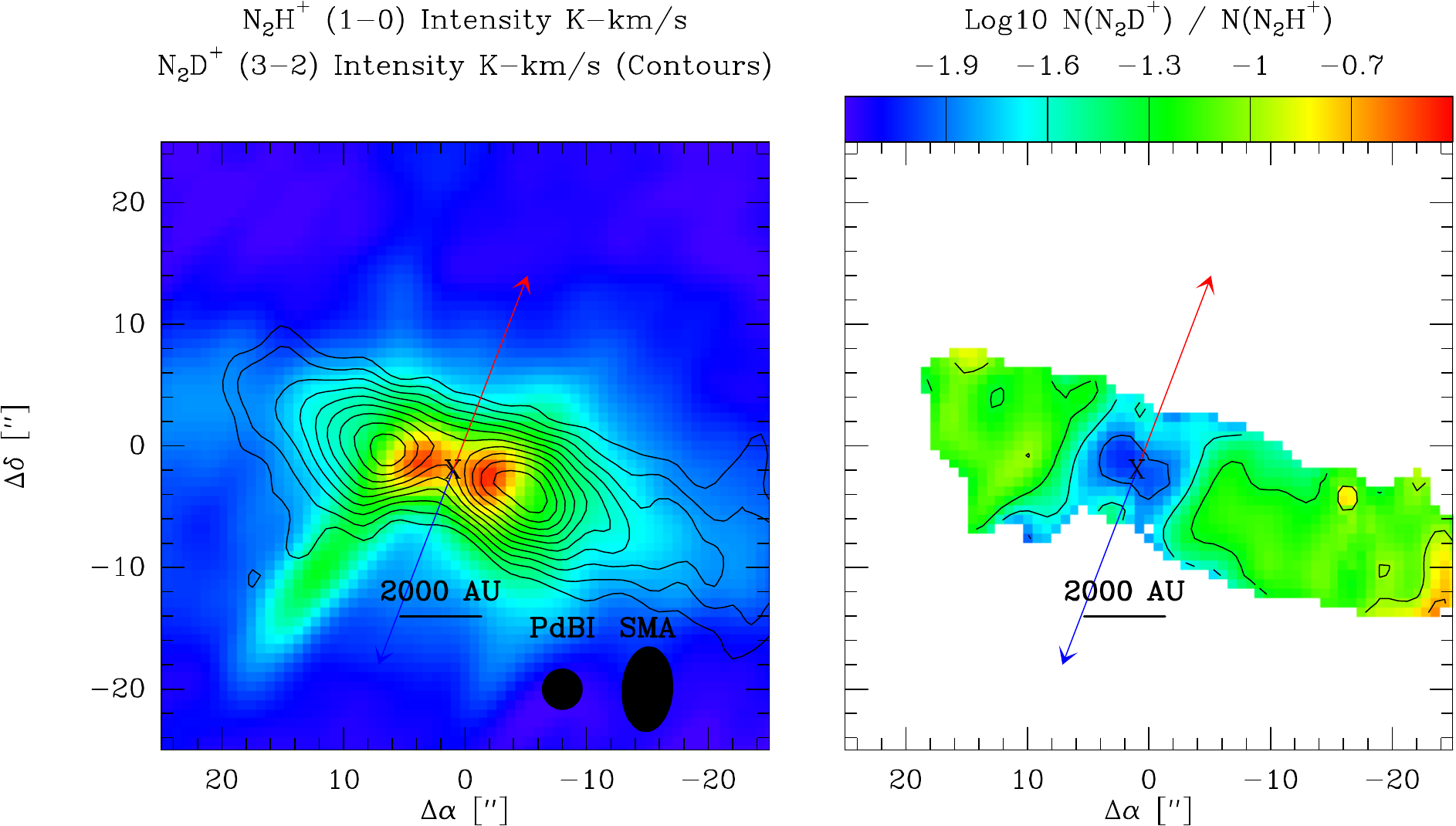}
\end{center}
\caption{Comparison of \nthp\ and \ntdp\ emission and column density ratio map. 
The \textit{left panel} shows the \ntdp\ integrated intensity (contours) overlaid on the 
PdBI \nthp\ integrated intensity (color scale). The \textit{right panel} shows the column density
ratio of \ntdp\ to \nthp. The color stretch in the left panel goes from 0 to 21 K \kms\ and the contours start 3$\sigma$
and increase with this interval ($\sigma$ = 0.07 K \kms).
}
\label{L1157-ratiomap}
\end{figure}

\begin{figure}
\begin{center}
\includegraphics[scale=0.45]{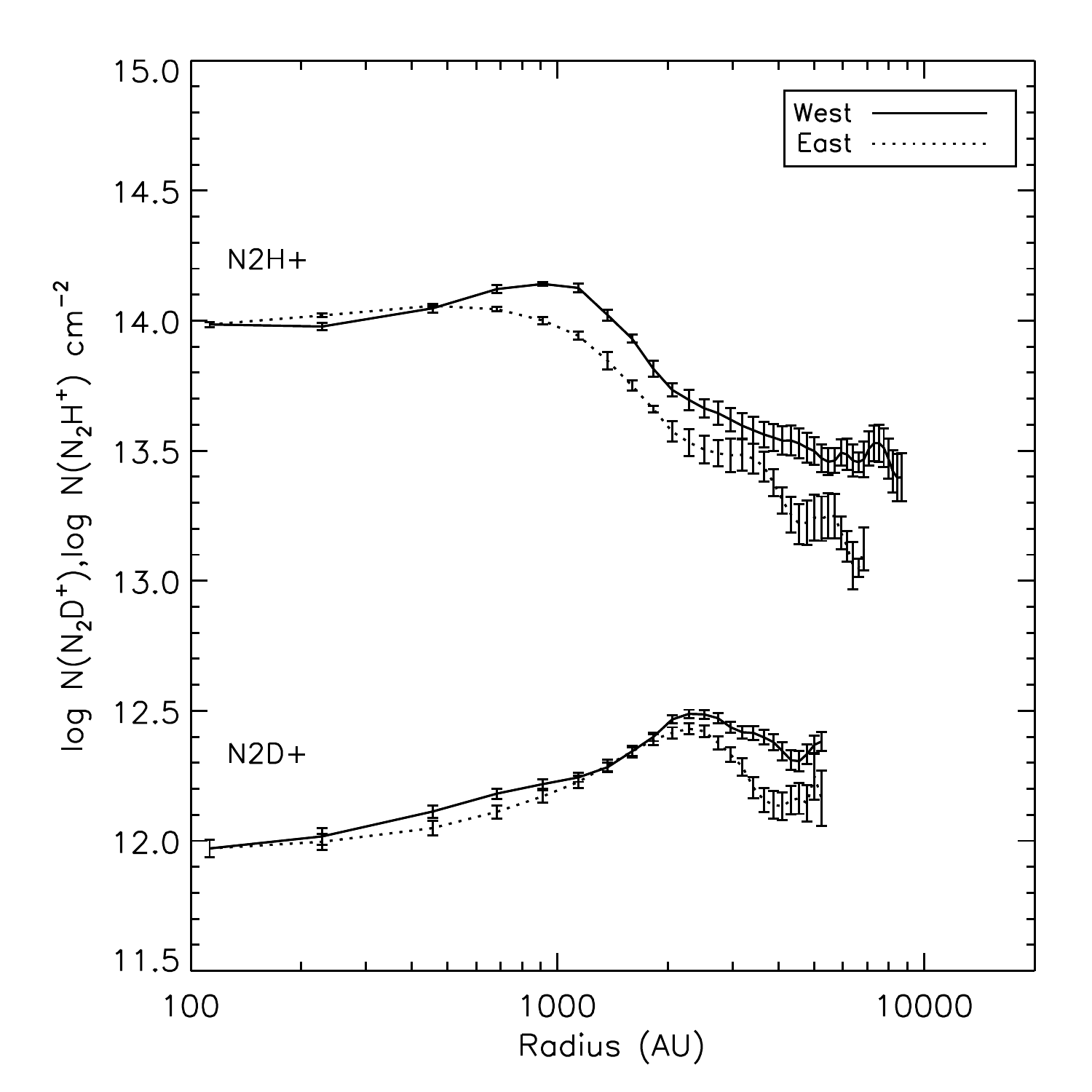}
\includegraphics[scale=0.45]{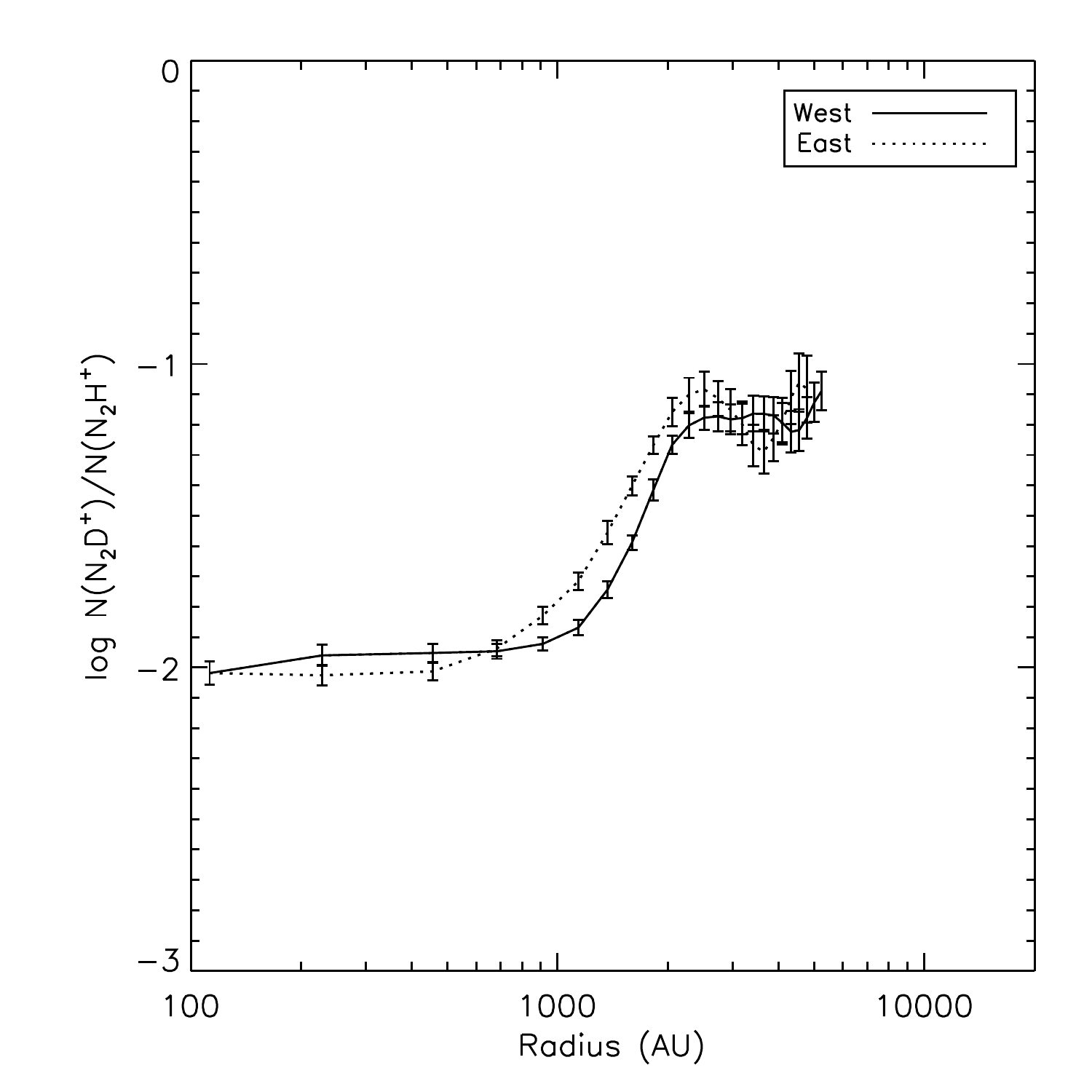}
\end{center}
\caption{\nthp\ and \ntdp\ column density and column density ratio plots. The \textit{left panel} 
shows the column density of \nthp\ and \ntdp\ derived from the integrated intensity maps. The right panel shows the
 \ntdp/\nthp\ column density ratio. Note the clear 1000 AU separation of \nthp\ and \ntdp\ column density peaks. The error bars on the \nthp\
data are the combination of noise, error in excitation temperature, and error in optical depth. The error bars on the
\ntdp\ data only include noise since the emission is optically thin and we adopt the excitation temperature of
the \nthp\ data. The solid lines are taken from
the west side of the envelope and the dotted lines are from the east side.
}
\label{L1157-abund}
\end{figure}

\begin{figure}
\begin{center}
\includegraphics[scale=0.7]{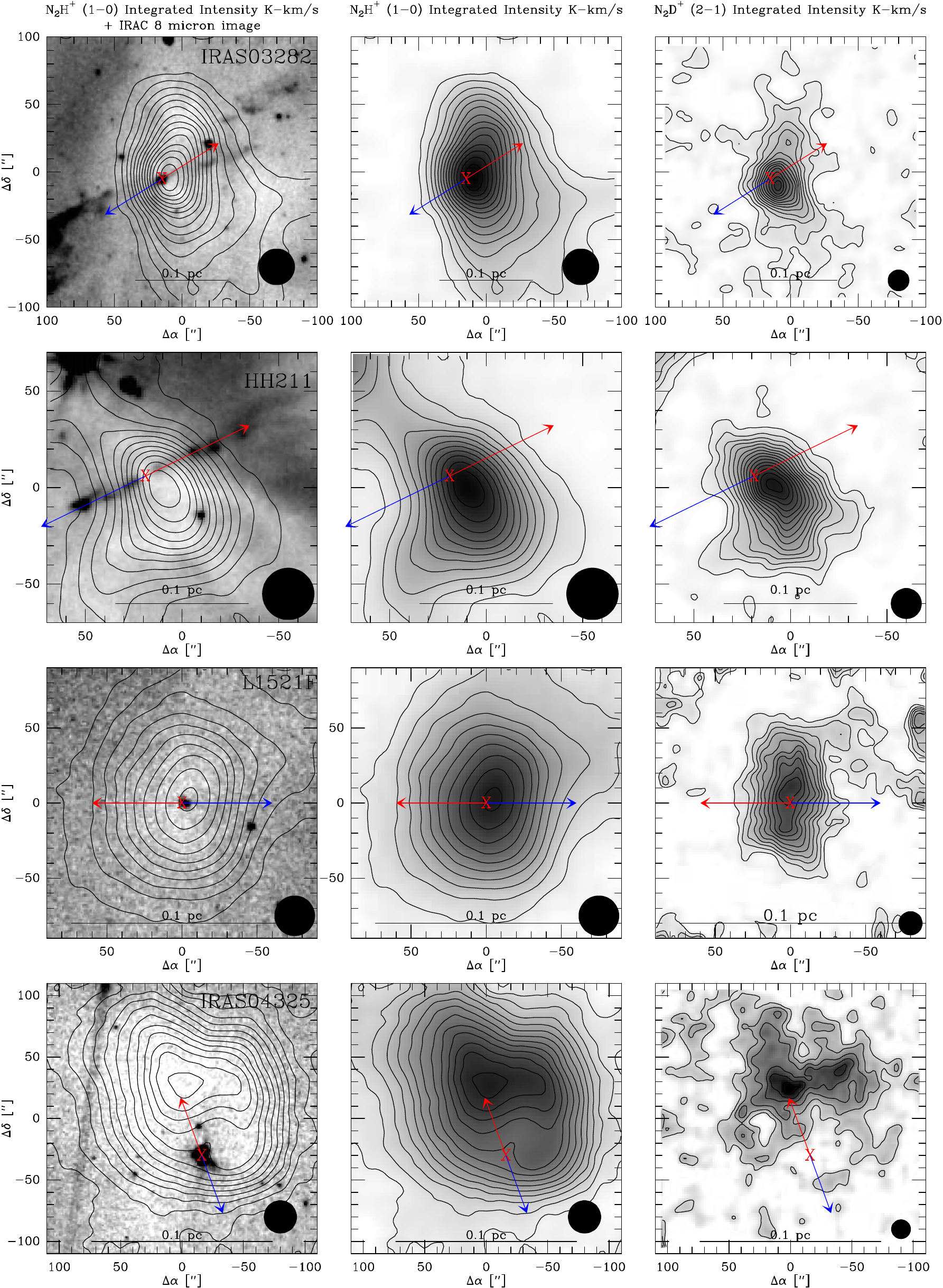}
\end{center}
\caption{Single-dish integrated intensity maps of \nthp\ ($J=1\rightarrow0$) and \ntdp\ ($J=2\rightarrow1$) 
for the protostars identified in each panel. The left panels show the \nthp\ ($J=1\rightarrow0$)
emission overlaid on the IRAC 8 \micron\ image, the middle panels show the \nthp\ ($J=1\rightarrow0$) as contours
and grayscale, and the right panels show the \ntdp\ ($J=2\rightarrow1$) emission as grayscale and contours. Eight of thirteen well-resolved
sources show spatial differentiation between the \nthp\ and \ntdp\ emission peaks; however, 
they are not as regular as in L1157. All \nthp\ contours start at 10$\sigma$ and increase by 10$\sigma$ and
all the \ntdp\ contours start at 3$\sigma$ and increase in 3$\sigma$ intervals; the values for $\sigma$ are given in Table 3 for
each source. The IRAM 30m beam for the \nthp\  ($J=1\rightarrow0$) data is 27\arcsec\ and 16\arcsec\ for the \ntdp\ ($J=2\rightarrow1$) data.
}
\end{figure}

\begin{figure}
\figurenum{6b}
\begin{center}
\includegraphics[scale=0.75]{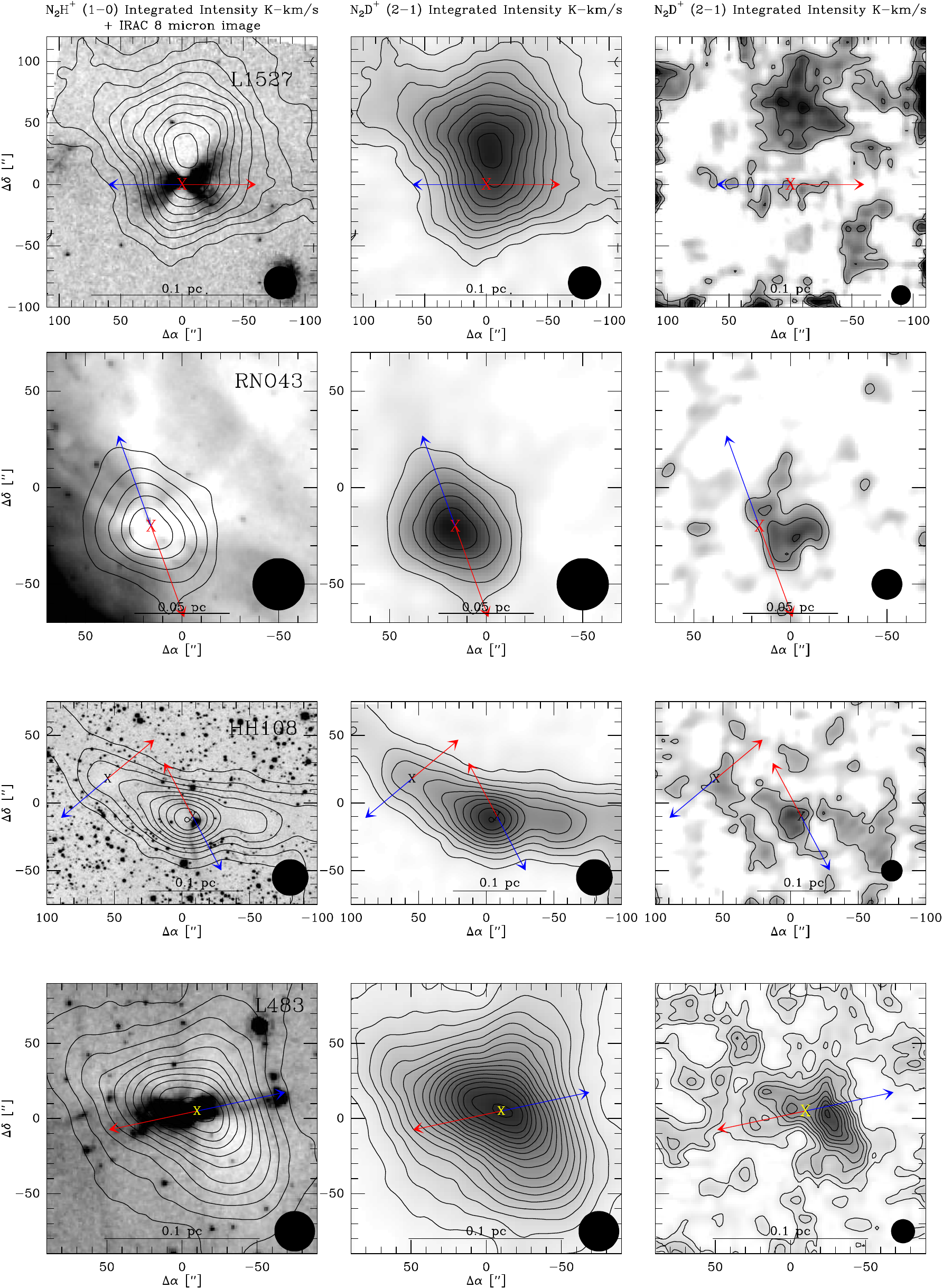}
\end{center}
\caption{}
\end{figure}

\begin{figure}
\figurenum{6c}
\begin{center}
\includegraphics[scale=0.75]{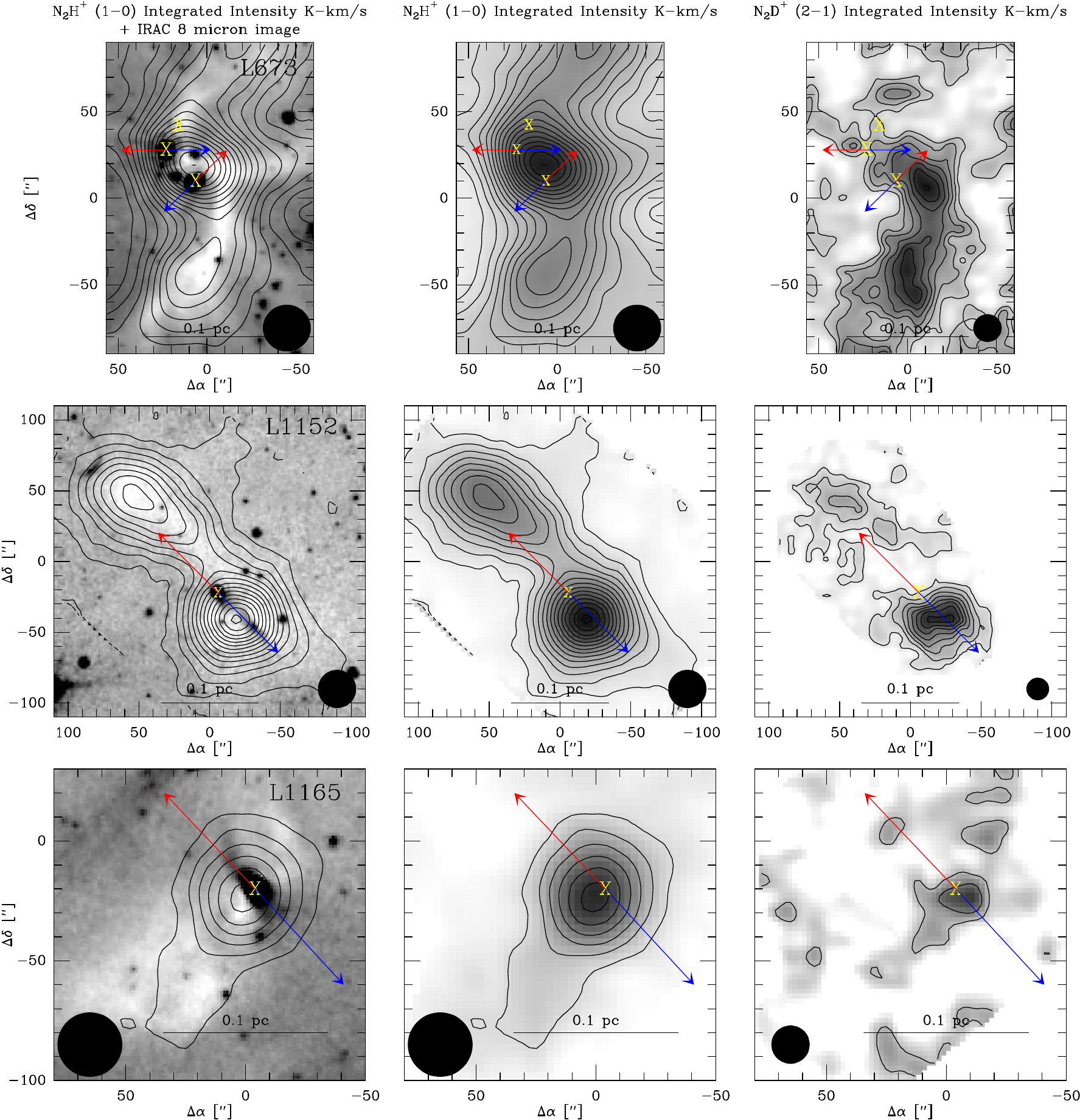}
\end{center}
\caption{}
\end{figure}

\begin{figure}
\begin{center}
\includegraphics[scale=0.7]{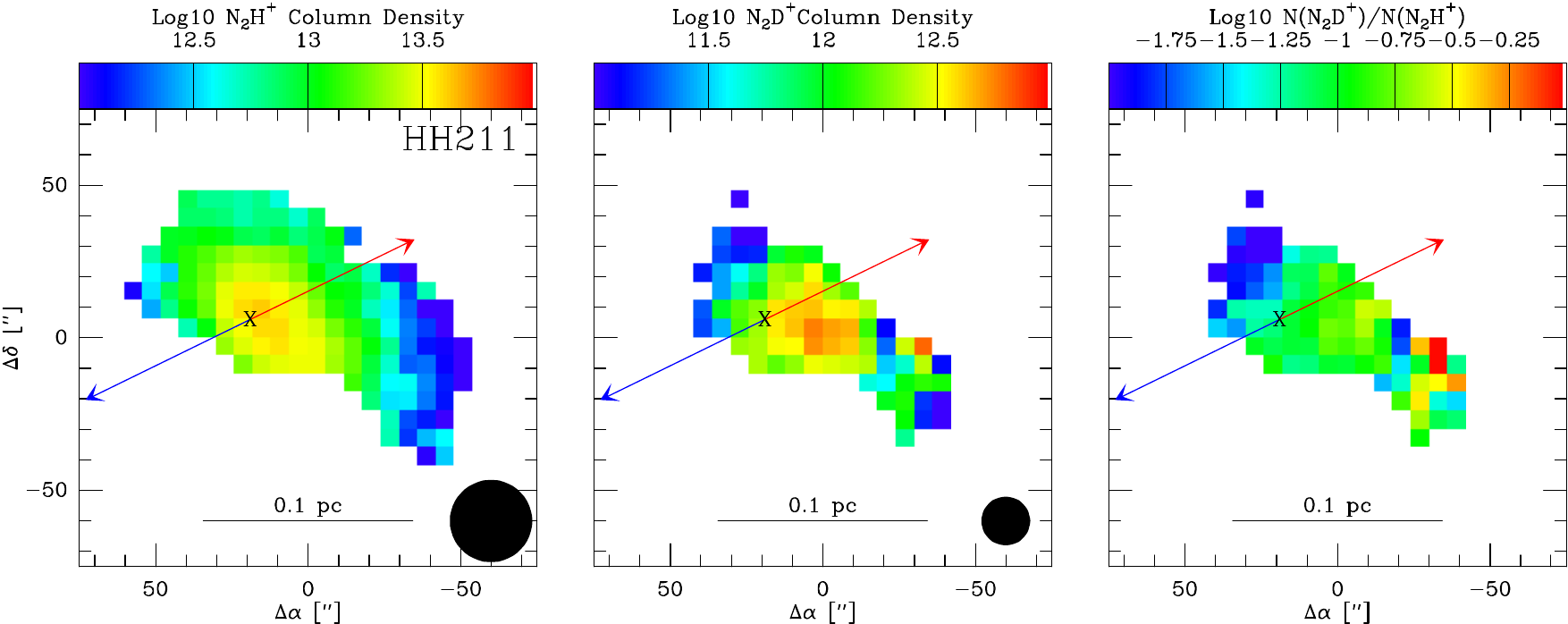}
\end{center}
\caption{Spatial distribution of \nthp\ column density (left panel), 
\ntdp\ (center panel), and their ratios (right panel) for HH211. Notice the column density
peak of \nthp\ at the location of the protostar, while the emission peak was offset in Figure 
6.}
\end{figure}

\begin{figure}
\begin{center}
\includegraphics[scale=0.7]{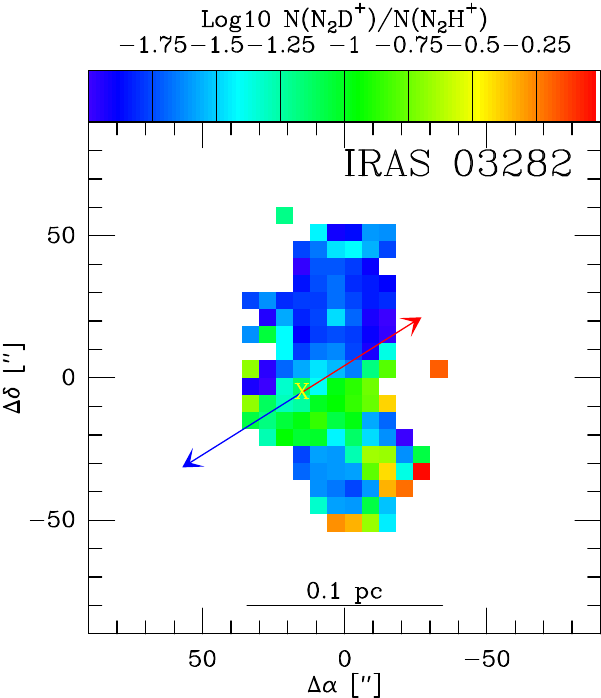}
\includegraphics[scale=0.7]{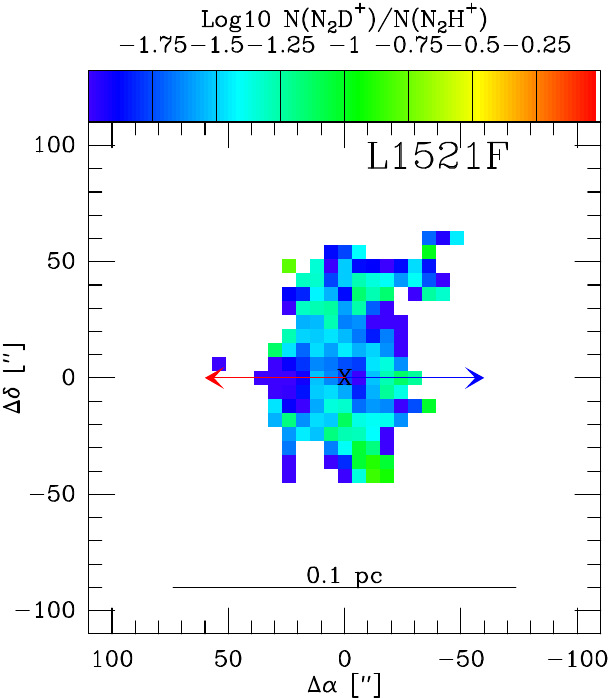}
\includegraphics[scale=0.7]{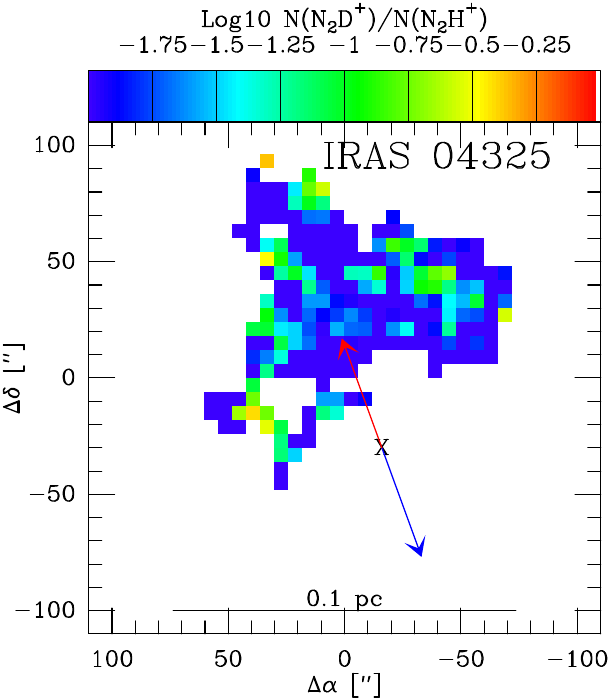}\\
\includegraphics[scale=0.7]{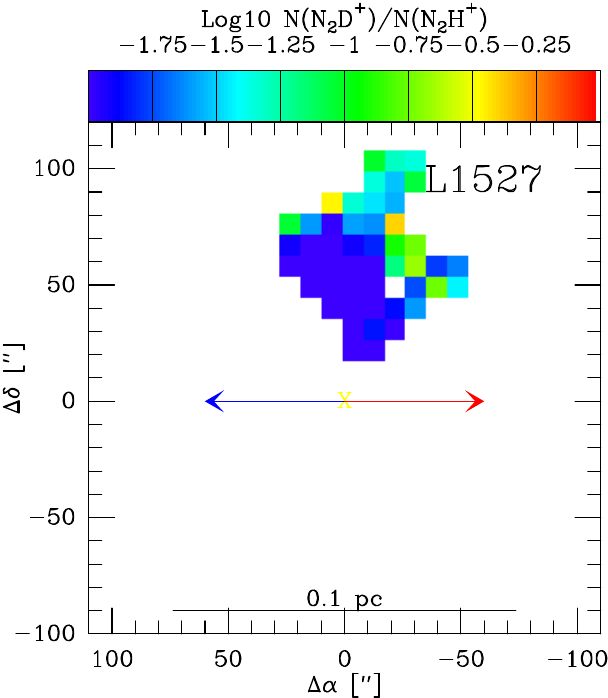}
\includegraphics[scale=0.7]{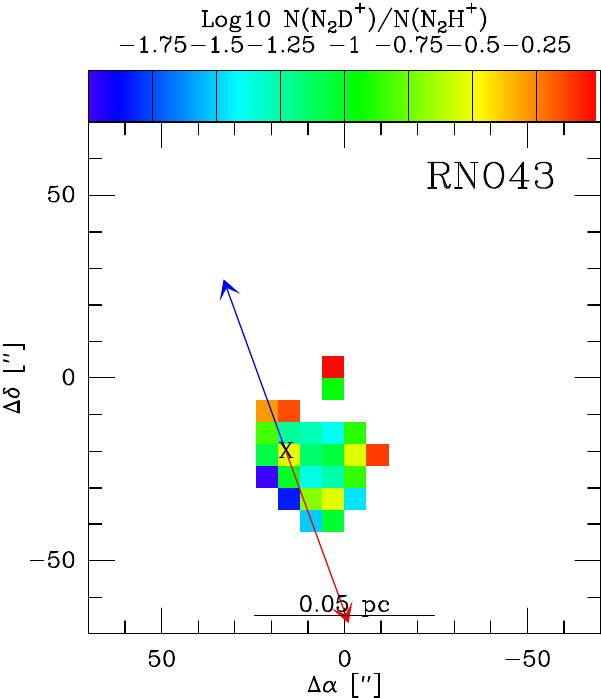}
\includegraphics[scale=0.7]{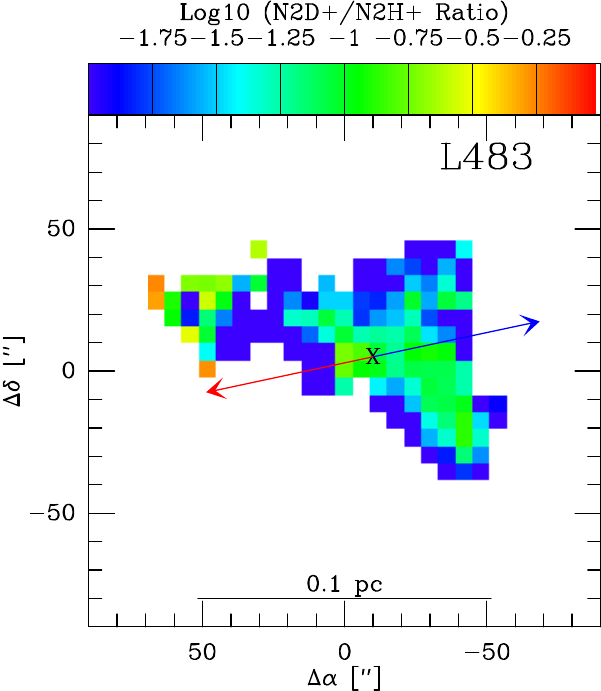}\\
\includegraphics[scale=0.7]{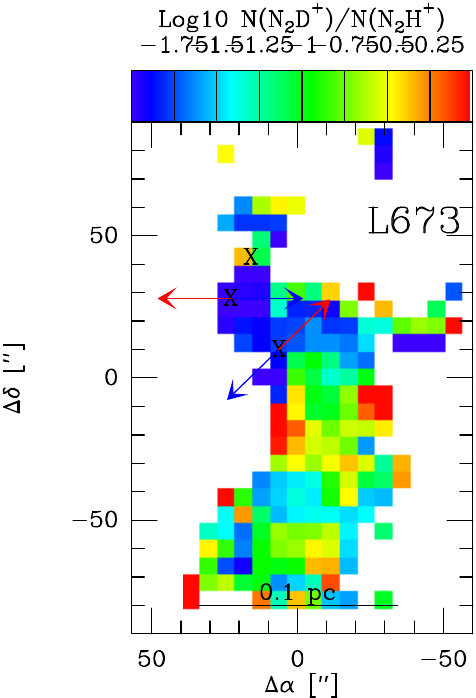}
\includegraphics[scale=0.7]{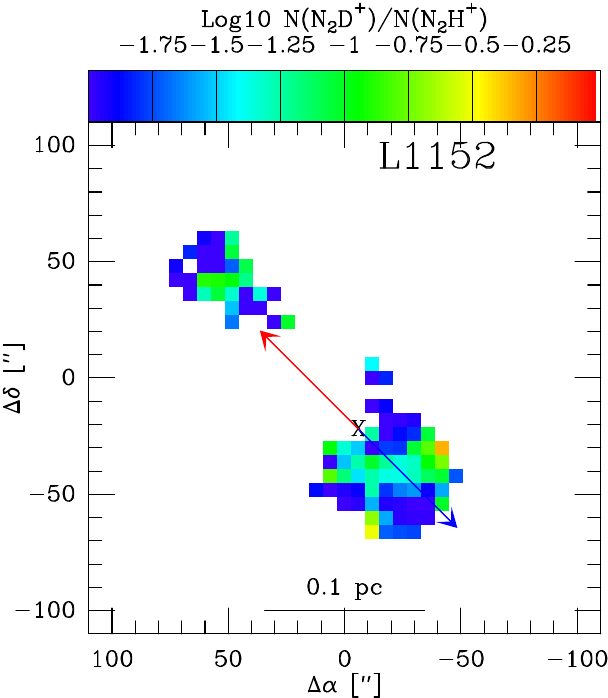}
\includegraphics[scale=0.7]{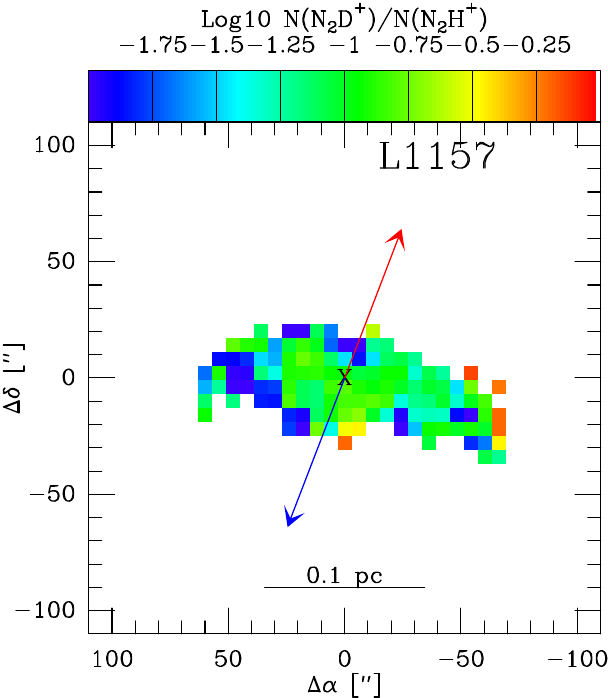}
\end{center}
\caption{Column density ratio maps for all sources except HH211 (Figure 7), L1165, and HH108.
}
\end{figure}

\clearpage

\begin{figure}
\begin{center}
\includegraphics[scale=0.45]{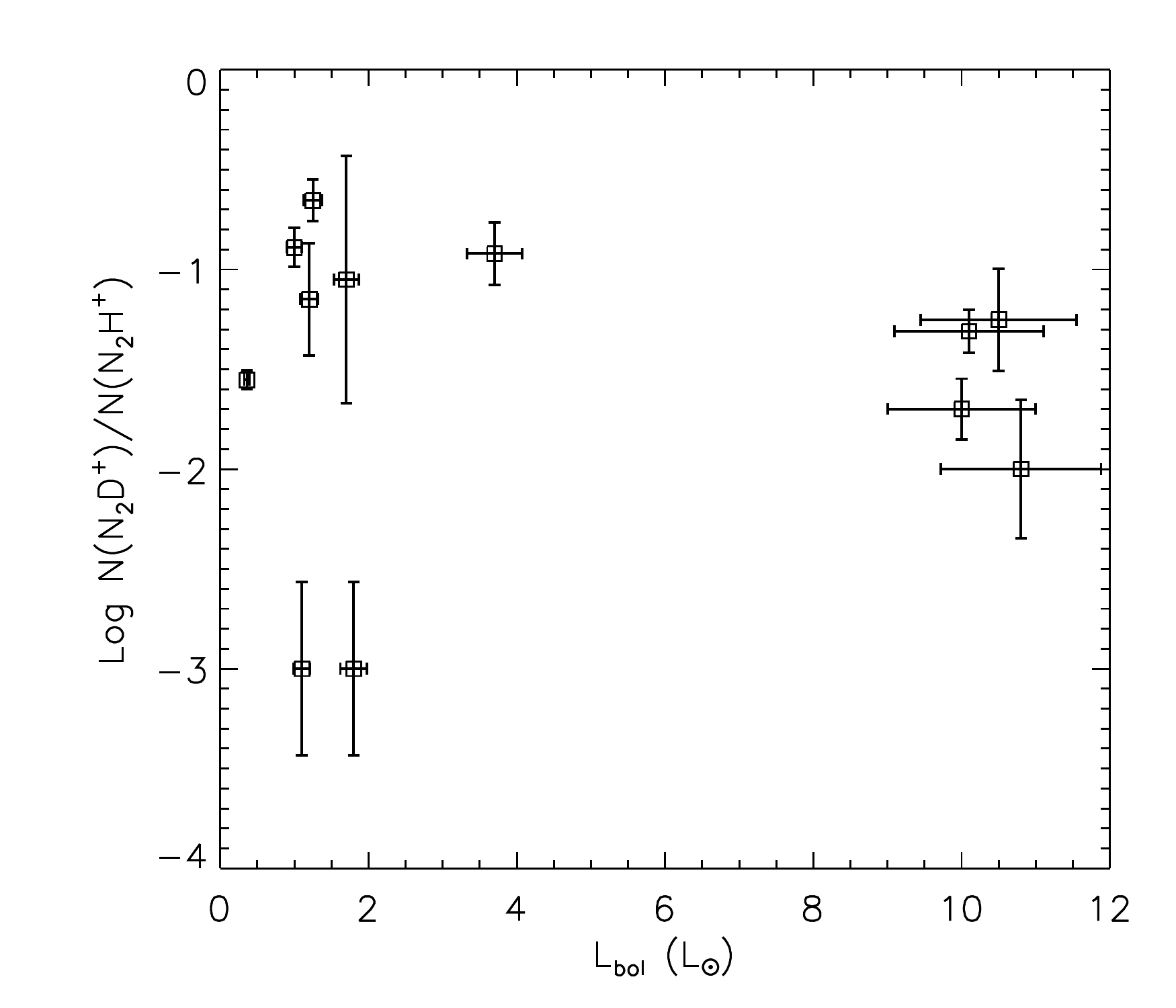}
\includegraphics[scale=0.45]{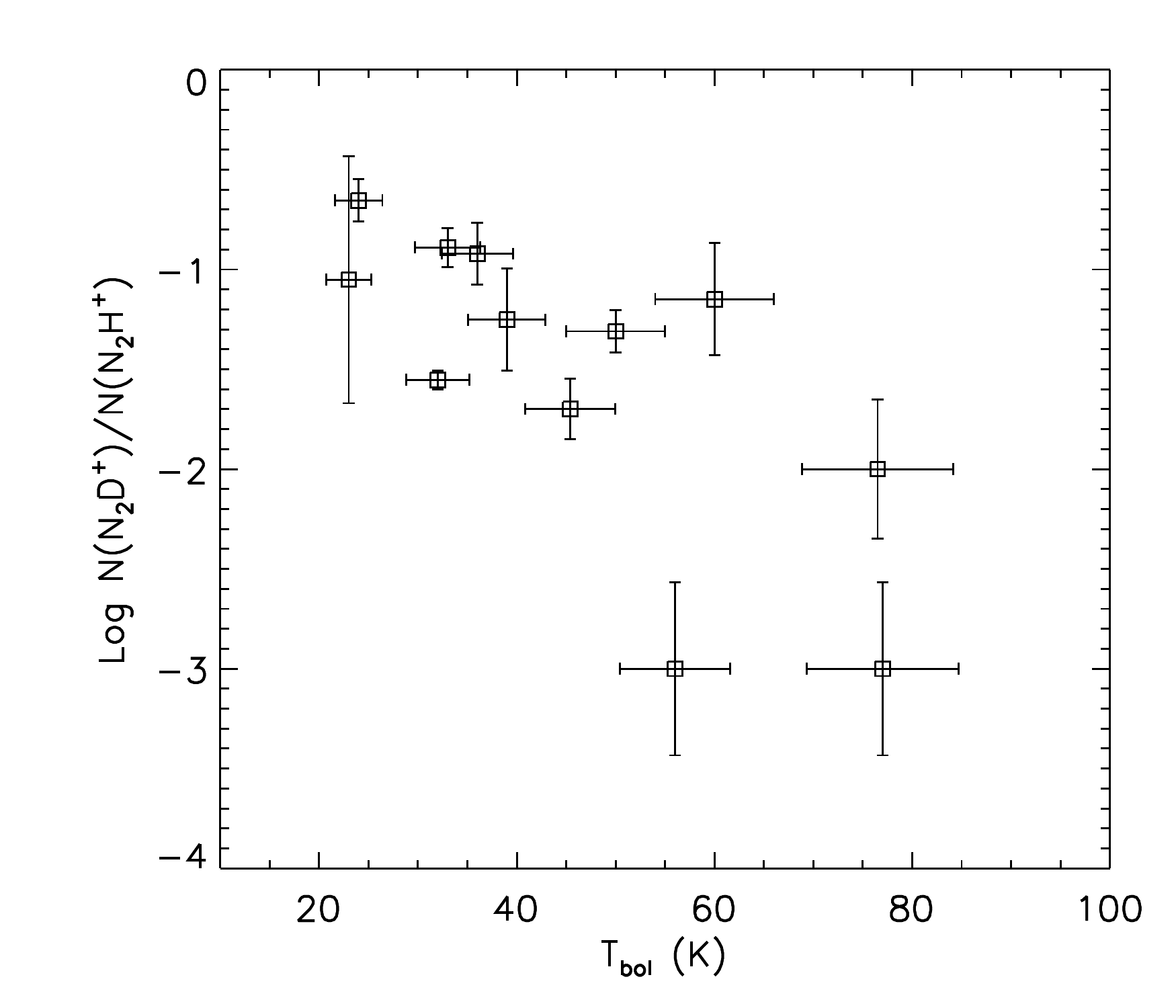}
\end{center}
\caption{\ntdp\ to \nthp\ ratio plotted versus bolometric luminosity (left panel) and bolometric temperature (right panel).
Higher luminosity sources tend to have a lower \ntdp/\nthp\ ratio, while lower luminosity sources have a range of deuteration.
There appears to be a trend in bolometric temperature versus  \ntdp/\nthp\ ratio with 
deuteration decreasing with evolution, consistent with \citep{emp2009}. 
}
\label{lboltbol}
\end{figure}

\begin{figure}
\begin{center}
\includegraphics[scale=0.85]{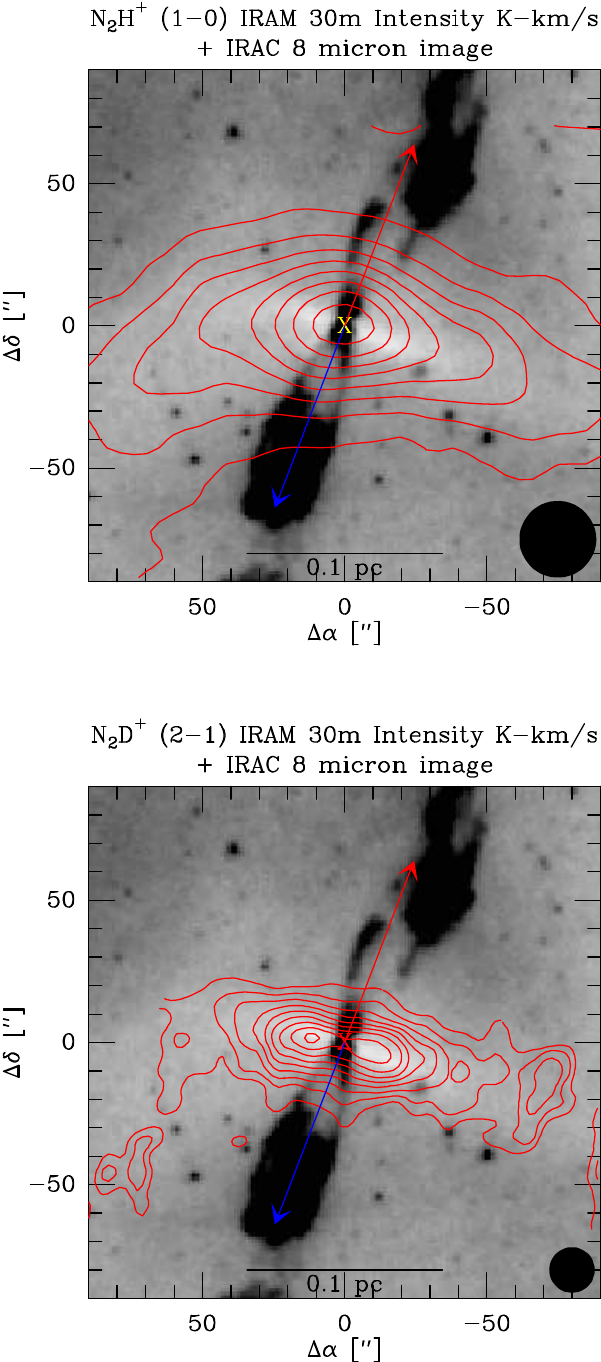}
\includegraphics[scale=0.85]{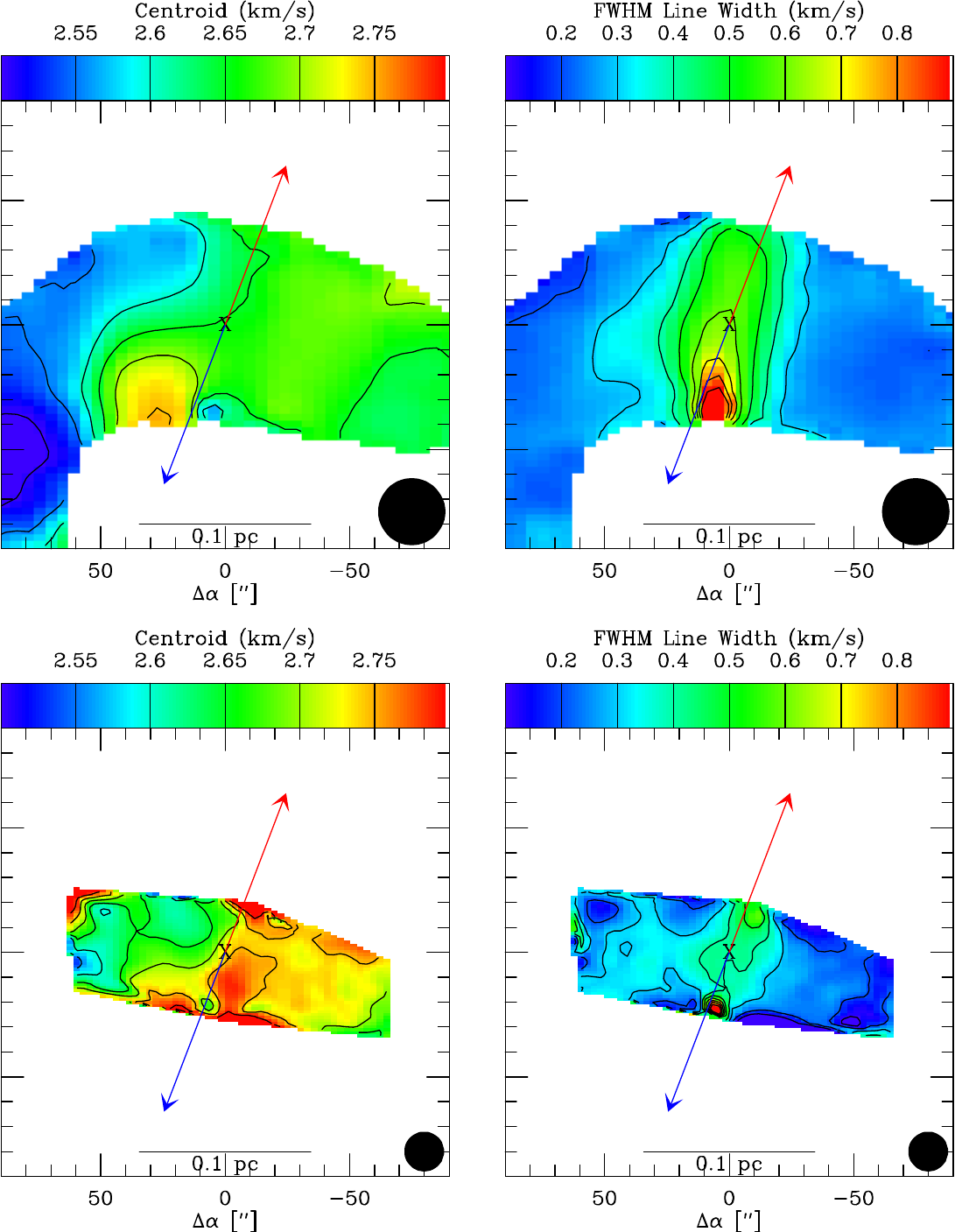}
\end{center}
\caption{Kinematic comparison of \nthp\ and \ntdp\ from the IRAM 30m data of L1157. The top row shows the IRAM 30m \nthp\ ($J=1\rightarrow0$) integrated intensity 
(\textit{left panel}), line-center velocity (\textit{middle panel}), and linewidth (\textit{right panel}). The bottom row
shows the same information as the top but for \ntdp\ ($J=2\rightarrow1$). The line-center velocities for \ntdp\ are about 0.1 \kms\
more redshifted on the west side of the envelope than on the east side and the linewidths of \ntdp\ are lower than the
\nthp\ linewidths toward the protostar.
}
\label{L1157-sd-kinematics}
\end{figure}

\begin{figure}
\begin{center}
\includegraphics[scale=0.85]{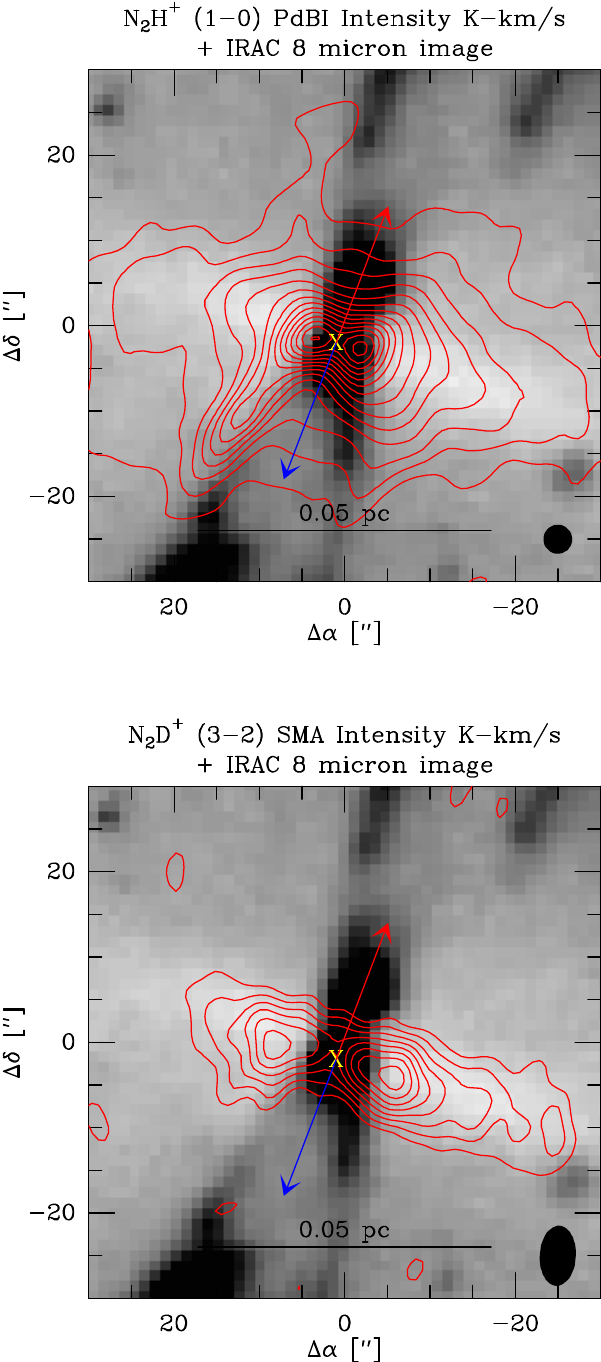}
\includegraphics[scale=0.85]{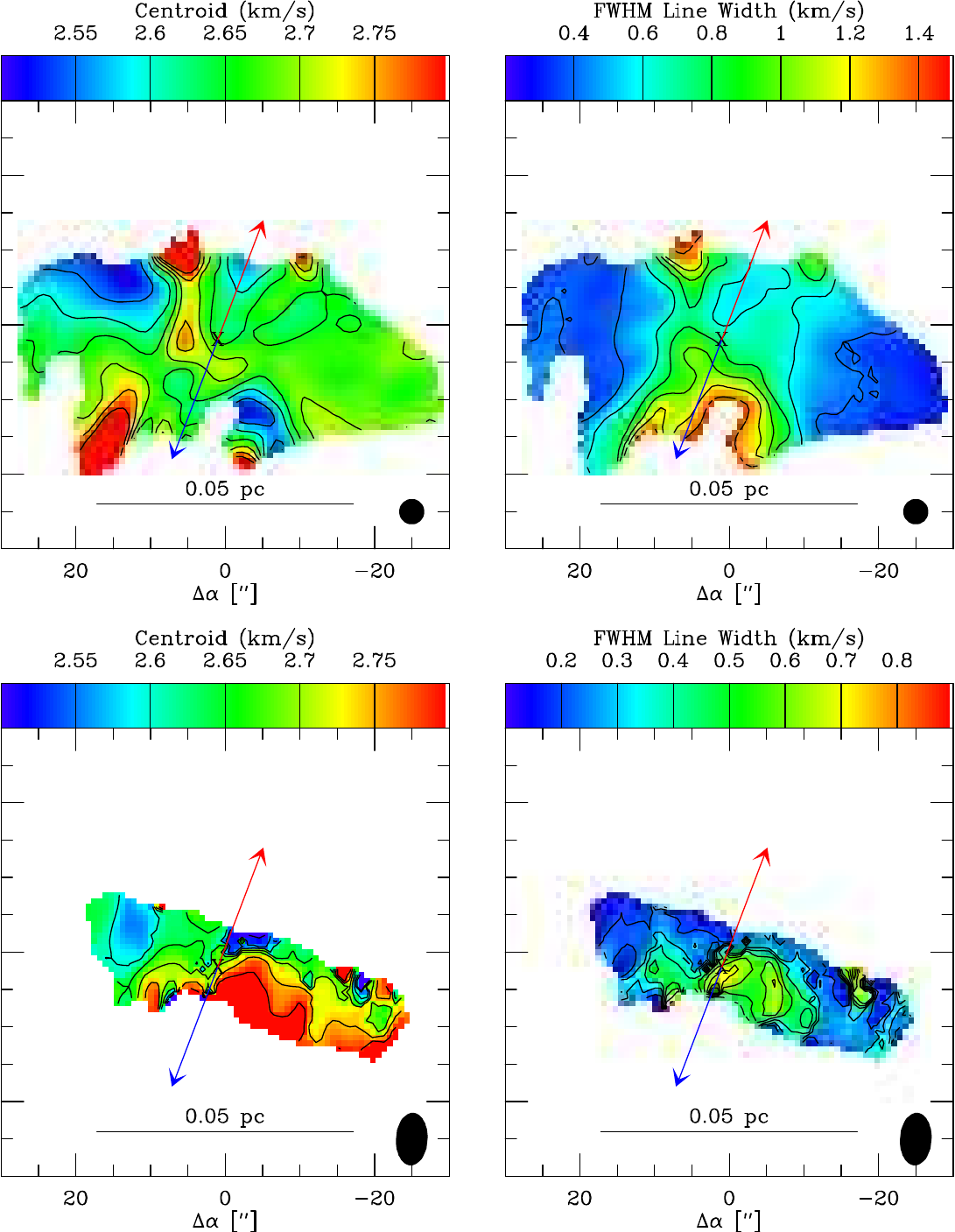}
\end{center}
\caption{Kinematic comparison of interferometric \nthp\ and \ntdp\ data of L1157. The top row shows the PdBI \nthp\ ($J=1\rightarrow0$) integrated intensity 
(\textit{left panel}), line-center velocity (\textit{middle panel}), and linewidth (\textit{right panel}). The bottom row
shows the same information as the top but for SMA \ntdp\ ($J=3\rightarrow2$). The velocity structure of \nthp\ in the line-center velocities
and linewidth is much more complex than \ntdp\ due to the outflow interaction with the envelope \citep{tobin2011}. The more redshifted
gas on the west side of the envelope also appears in the \ntdp\ ($J=3\rightarrow2$) which is also present in the single-dish \ntdp.
}
\label{L1157-int-kinematics}
\end{figure}

\begin{figure}
\begin{center}
\includegraphics[scale=0.5]{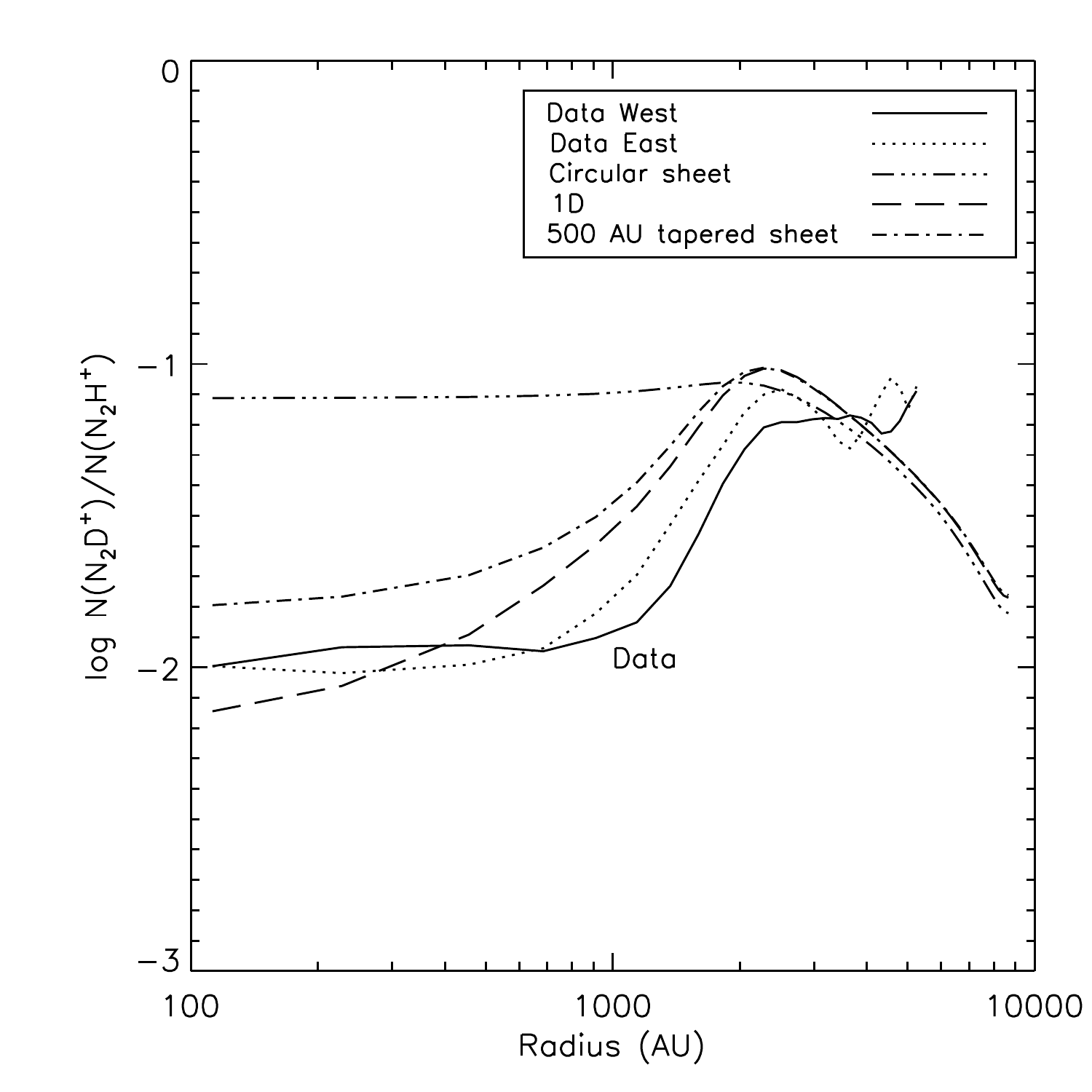}
\includegraphics[scale=0.5]{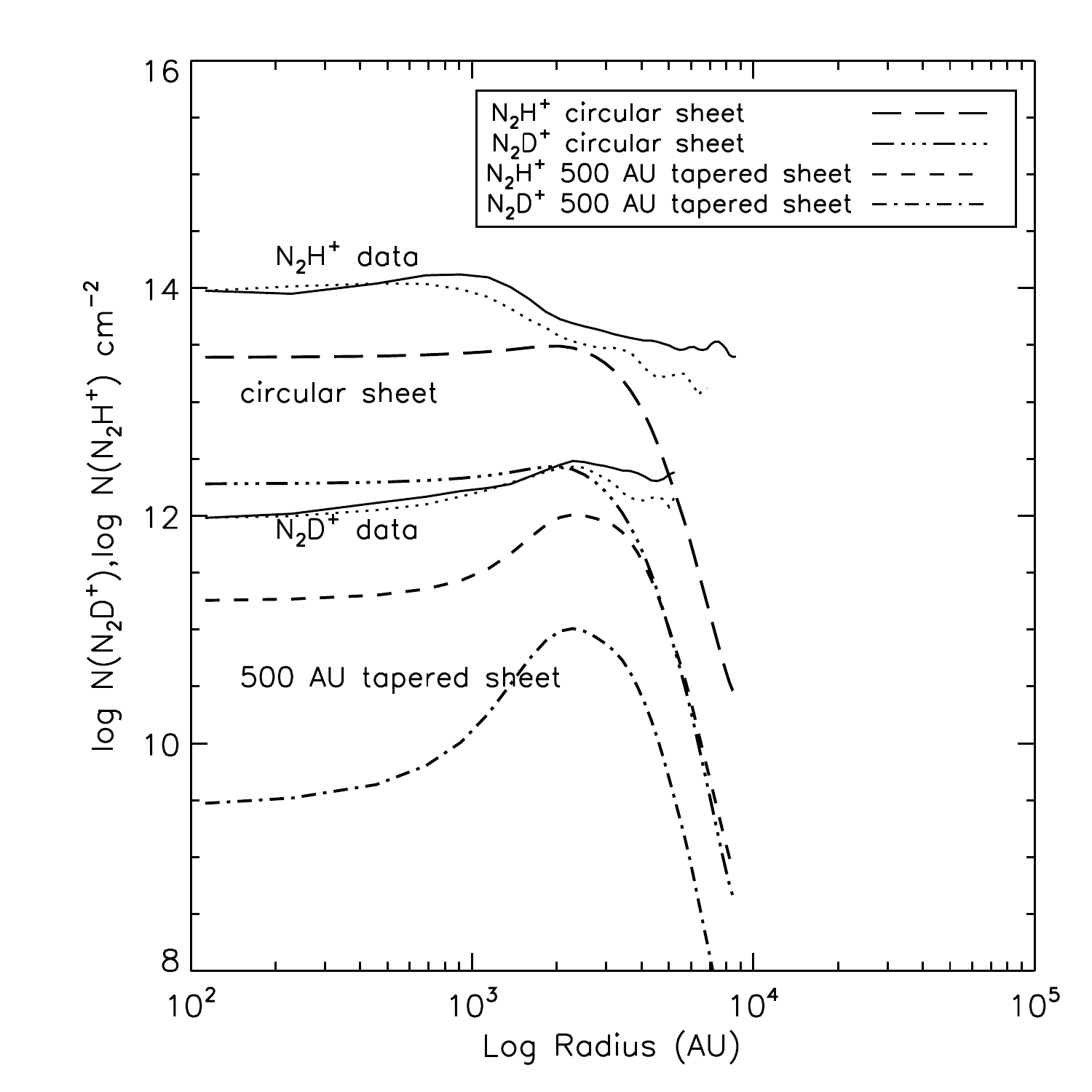}
\end{center}
\caption{Abundance ratio and integrated column densities of \nthp\ and \ntdp\ compared to models. 
The abundance ratio of the 1D model and tapered sheet (filament) resemble the abundance ratio
of the data; however, the absolute column densities do not match the data.
The \textit{left} panel shows the abundance ratio of \ntdp\ to \nthp\ from the 1D chemical model calculation (dashed-line), 
integrated through a circular sheet
with a radius of 15000 AU (triple dot-dashed line) and an approximated filamentary envelope, produced
by tapering the circular sheet by a Gaussian with $\sigma$ = 500 AU (single dot-dashed line).
 The data are shown as the solid and dotted lines from the east and west
sides of the envelope respectively. The \textit{right} panel shows the column density of \nthp\ and \ntdp\ 
derived from integrated maps, assuming that \ntdp\ has the same excitation temperature
as \nthp. The data are shown as the solid and dotted lines from the east and west
sides of the envelope respectively and the model data are shown as the dashed or dot-dashed lines.
Note the coincidence of column density peaks in the models for \ntdp; moreover, the peaks
are most apparent in the filament model rather than the sheet.
}
\label{L1157-abund-model}
\end{figure}


\begin{figure}
\begin{center}
\includegraphics[scale=0.75]{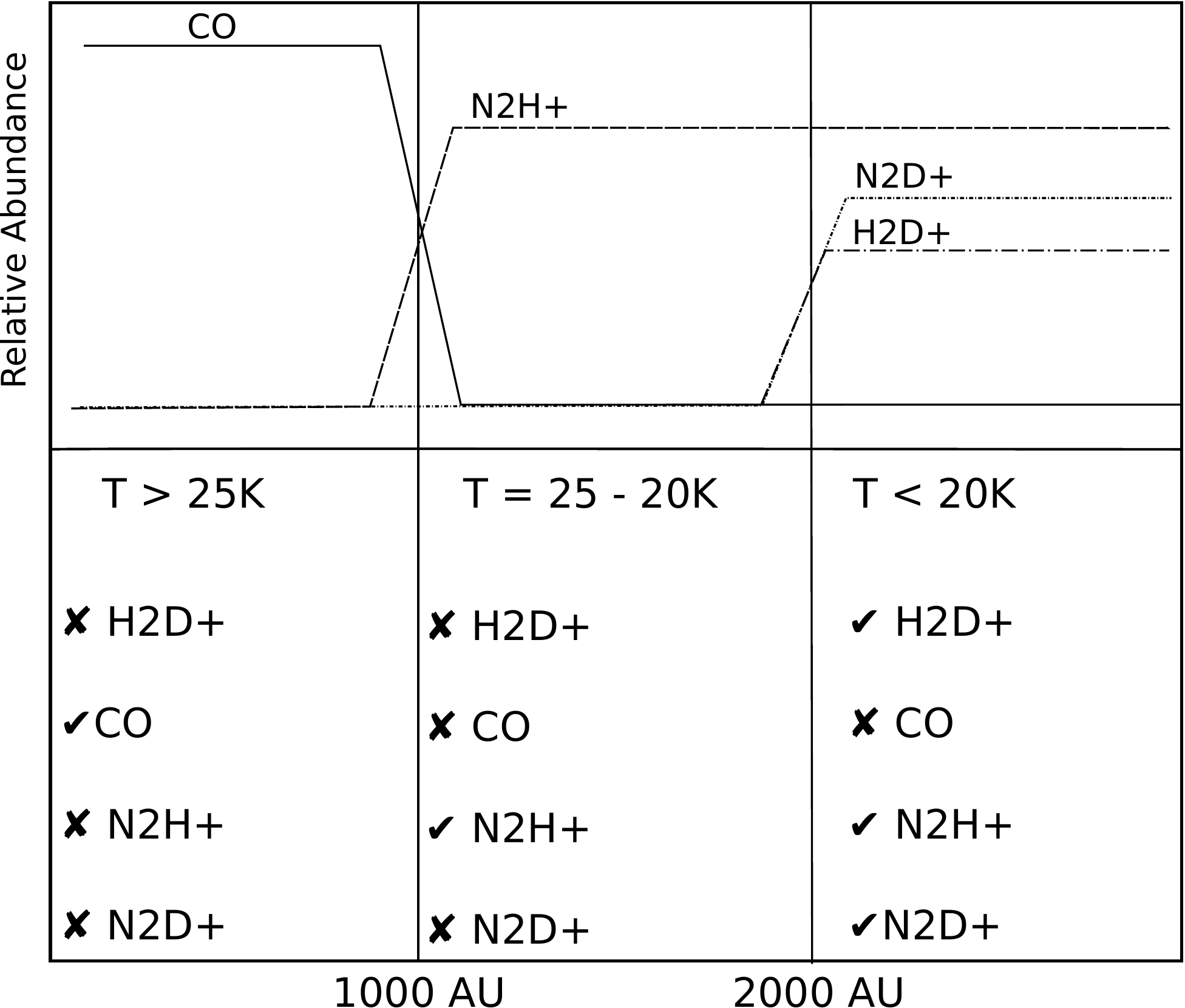}

\end{center}
\caption{Schematic diagram for spatial distribution of \nthp\ and \ntdp\ in L1157. The upper panel
shows the suggested relative abundances of each species as a function of radius and temperature and the
bottom panel identifies the temperature range and which species are present in the gas phase for the
case of increased CO evaporation temperature. Inside of 1000 AU, CO is
in the gas phase, evaporated off the dust grains, and destroys the \nthp. At radii between 1000 AU and 2000 AU
the CO may remain frozen on the dust grains at the expected temperatures of 20 to 25 K; this enables \nthp\
to be abundant, but \htdp\ is destroyed due to the back reaction operating (Equation 1) at these temperatures, keeping
\ntdp\ abundances low. At radii greater than 2000 AU, the temperature is below 20 K and \htdp\ will be present, enabling \ntdp\
to form and have its abundance peaks at this radius. If there is an H$_2$ OPR gradient rather than an increased CO
evaporation temperature, \ntdp\ and \htdp\ may only be present at T $<$ 15 K and CO could then evaporate at T $>$ 20 K
and still show the same abundance offset.
}
\label{n2dp-schematic}
\end{figure}

\begin{deluxetable}{lllllllllll}
\rotate
\tablewidth{0pt}
\tabletypesize{\scriptsize}
\tablecaption{Source Properties}
\tablehead{
  \colhead{Source}  & \colhead{RA} & \colhead{Dec}      & \colhead{Distance}    &  \colhead{Mass$_{8\mu m}$} & \colhead{Mass$_{submm}$\tablenotemark{*}}   & \colhead{L$_{bol}$}     & \colhead{T$_{bol}$}  & \colhead{Outflow PA} & \colhead{References}\\
                    & \colhead{(J2000)} &  \colhead{(J2000)}     & \colhead{(pc)}          &  \colhead{($M_{\sun}$)}     & \colhead{($M_{\sun}$)}                      & \colhead{($L_{\sun}$)}  & \colhead{(K)}        & \colhead{(\degr)} & \colhead{(Distance, M$_{ref}$, L$_{bol}$,)}  \\
                    &                   &     &                                           &  \colhead{(r$<$0.05pc)}     &                                    &                                   &                            &                   & \colhead{(T$_{bol}$, Outflow PA)}
}
\startdata
IRAS 03282+3035 & 03:31:21.10 & +30:45:30.2 & 230   & 2.4 & 2.2          & 1.2   & 33      & 122 & 23, 4, 3, 3, 26\\
HH211           & 03:43:56.78 & +32:00:49.8 & 230  & 1.1 &   1.5        & 3.02  & 24       & 116 & 23, 3, 3, 3, 6  \\
L1521F          & 04:28:39.03 & +26:51:35.0 & 140 & 2.3 &  1.0         & 0.35  & $\sim$20  & 270 & 20, 10, 8, 1, 7 \\
IRAS 04325+2402 & 04:35:35.39 & +24:08:19.0 & 140   &  3.9   & -          &  0.97 & 73       & 200 & 20, 1, 15, 15, 16 \\
L1527           & 04:39:53.86 & +26:03:09.5 & 140  & 0.8  & 2.4         & 1.9   & 56       & 90 & 20, 4, 9, 15, 31\\
RNO43           & 05:32:19.39 & +12:49:40.8 & 460 & 2.8  & 2.6          & 12.5 & 56        & 20 & 21, 10, 2, 15, 21 \& 1\\
L483            & 18:17:29.93 & -04:39:39.6 & 200  & 3.5  & 1.8          & 11.5 & $<$54    & 282 & 25, 4, 3, 15, 30 \\
HH108 IRS        & 18:35:42.14 & -00:33:18.5 & 300 &  -   & 4.5    & $\sim$8.0 & 28  & 208  & 17, 14, 14, 1, 1\\
HH108 MMS        & 18:35:46.46 & -00:32:51.2 & 300 &  -   & 3.6    & 0.7   & 18     & 130   & 17, 17, 1, 1, 1\\
L673-SMM2       & 19:20:25.96 & +11:19:52.9 & 300 & 1.0  & 0.35        &  2.8  & -         & 270, 135 & 24, -, -, 12, 1 \& 12 \\
L1152           & 20:35:46.22 & +67:53:01.9 & 300 & 3.4  & 12.0        & 1.0   & 33        & 225 & 5, 10, 3, 1, 11 \\
L1157           & 20:39:06.25 & +68:02:15.9 & 300 & 2.6  & 2.2        & 3.0  & 29         & 150 & 5, 4, 3, 15, 27\\
L1165           & 22:06:50.46 & +59:02:45.9  & 300 & 1.1  & 0.32        & 13.9 & 46        & 225 & 5,  1, 12, 1, 12\\
 
\enddata
\tablecomments{ Properties of sources observed in the single-dish and/or interferometric sample. The 8\mum\ extinction
masses are taken within 0.05 pc of the protostar and note that some of the masses have been rescaled to account 
for a different distance estimate as compared to Paper I. Positions are reflect the coordinates of the 24\mum\ 
point source from \textit{Spitzer} data or the 3mm continuum continuum position for protostars observed with CARMA.
The Outflow position axes (PA) are not well constrained since the outflows are known to precess they can have fairly large
angular width; a conservative estimate of uncertainty would be $\pm$10\degr. 
References: (1) This work, (2) \citet{tobin2010a}, (3) \citet{enoch2009}, 
 (4) \citet{shirley2000}, (5) \citet{kirk2009}, (6) \citet{lee2009}, (7) \citet{bourke2006}
 (8) \citet{terebey2009}, (9) \citet{tobin2008}, (10) \citet{young2006},
 (3) \citet{visser2002}, (11) \citet{chapman2009}, (14) \citet{enoch2007},
 (15) \citet{froebrich2005}, (16) \citet{hartmann1999}, (17) \citet{chini2001},
 (20) \citet{loinard2007}, (21) \citet{bence1996},
(23) \citet{hirota2011}, (24) \citet{herbig1983}, (25) \citet{jorgensen2004}, (26) \citet{arce2006},
(27) \citet{gueth1996},  (30) \citet{fuller1995},
(31) \citet{hoger1998}.} 
\tablenotetext{*}{Mass computed from sub/millimeter emission assuming an isothermal temperature.}

\end{deluxetable}

\begin{deluxetable}{lllccc}
\tabletypesize{\scriptsize}
\tablewidth{0pt}
\tablecaption{Observed Species}
\tablehead{   \colhead{Transition} & \colhead{Frequency} & \colhead{Observatory} & \colhead{Beam width} & \colhead{F$_{eff}$} & \colhead{B$_{eff}$}\\
  & \colhead{(GHz)} & & \colhead{($^{\prime\prime}$)} &  & }
\startdata
\nthp\ ($J=1\rightarrow0$) & 93.1737637 & IRAM 30m & 26 & 0.95 & 0.81 \\
             &            & PdBI D, C-array  &  3.4 x 3.3\\

\nthp\ ($J=3\rightarrow2$) & 279.511832 & IRAM 30m  & 9  & 0.88 & 0.53 \\
\ntdp\ ($J=2\rightarrow1$) & 154.217206 & IRAM 30m & 16  & 0.93 & 0.74 \\
\ntdp\ ($J=3\rightarrow2$) & 231.321966 & IRAM 30m & 10   & 0.94 & 0.63 \\
             &            & SMA sub-compact & 7.1 x 4.3\\
\enddata
\tablecomments{The single-dish data were rescaled to main beam brightness 
temperatures (T$_{mb}$) where T$_{mb}$ = F$_{eff}$T$_{A}^*$/B$_{eff}$. F$_{eff}$ is
the forward efficiency, accounting for losses by the secondary support structure, and B$_{eff}$ is the main beam
efficiency, the ratio of the main beam solid angle over the entire antenna 
pattern (http://www.iram.es/IRAMES/mainWiki/Iram30mEfficiencies).}
\end{deluxetable}

\begin{deluxetable}{llllccl}
\tablewidth{0pt}
\tabletypesize{\scriptsize}
\tablecaption{IRAM 30m Observations}
\tablehead{
  \colhead{Source} & \colhead{RA}      & \colhead{Dec}      & \colhead{Date}   & \colhead{T$_{sys}$\tablenotemark{\dagger}} & \colhead{Channel Width\tablenotemark{\dagger}} & \colhead{$\sigma_{I}$\tablenotemark{\dagger}}\\
                   & \colhead{(J2000)} &  \colhead{(J2000)} & \colhead{(UT)}       & \colhead{(K)}                           & \colhead{(kHz)}& \colhead{(K \kms)}\\
}
\startdata
IRAS 03282$+$3035 & 03:31:19.8  & +30:45:37         & 25 Oct 2009        & 115, 130 & 20, 40 & 0.04, 0.035 \\
HH211-mm          & 03:43:55.3  & +32:00:46         & 24 Oct 2009        & 110, 120 & 20, 40 & 0.05, 0.04\\
L1521F            & 04:28:39.0  & +26:51:37         & 24 Oct 2009        & 102, 110 & 20, 40 & 0.04, 0.025\\
IRAS04325+2402    & 04:35:36.6  & +24:08:54         & 25 Oct 2009        & 105, 120 & 20, 40 & 0.04, 0.035\\
L1527             & 04:39:54.0  & +26:03:22         & 26 Oct 2009        & 100, 130 & 20, 40 & 0.03, 0.015\\
RNO43             & 05:32:18.2  & +12:50:01         & 24 Oct 2009        & 115, 135 & 20, 40 & 0.04, 0.03\\
L483              & 18:17:30.3  & -04:39:40.6       & 25 Oct 2009        & 135, 180 & 20, 40 & 0.06, 0.05\\
HH108             & 18:35:42.7  & -00:33:07         & 23 Oct 2009        & 140, 230 & 20, 40 & 0.05, 0.06\\
L673-SMM2         & 19:20:25.4  & +11:19:44         & 23/24 Oct 2009     & 140, 210 & 20, 40 & 0.04, 0.04\\
L1152             & 20:35:47.17 & +67:53:22         & 24 Oct 2009        & 120, 150 & 20, 40 & 0.04, 0.035\\
L1157             & 20:39:05.6  & +68:02:13.2       & 25 Oct 2009        & 110, 115, 320, 290  & 20, 20, 40, 40 & 0.027, 0.05, 0.12, 0.14\\
L1165             & 22:06:50.8  & +59:03:06.5       & 25 Oct 2009        & 110, 120 &  20, 40 & 0.04, 0.035\\

\enddata
\tablecomments{Observations of protostellar envelopes taken with the IRAM 30m telescope. The positions reflect the map
center and not necessarily the protostar position.} 
\tablenotetext{\dagger}{System temperatures, channel widths, and intensity uncertainties
 are given for 93 GHz and 154 GHz respectively. For L1157, values for 231 GHz and 279 GHz are also given.}
\end{deluxetable}

\begin{deluxetable}{llllllllllll}
\tabletypesize{\scriptsize}
\tablewidth{0pt}
\rotate
\tablecaption{\nthp\ and \ntdp\ Line Parameters}
\tablehead{   \colhead{Source} 
& \colhead{W} 
& \colhead{V$_{lsr}$}
& \colhead{$\Delta$V} 
& \colhead{$T_{ex}$}
& \colhead{$\tau$}
& \colhead{W} 
& \colhead{V$_{lsr}$} 
& \colhead{$\Delta$V} 
& \colhead{$T_{ex}$}
& \colhead{$\tau$}
\\
& \colhead{\nthp\ ($J=1\rightarrow0$)} 
& \colhead{}
& \colhead{} 
& \colhead{}
& \colhead{}
& \colhead{\ntdp\ ($J=2\rightarrow1$)} 
& \colhead{} 
& \colhead{} 
& \colhead{}
& \colhead{}
\\
   & \colhead{(K \kms)} & \colhead{(\kms)} & \colhead{(\kms)} & \colhead{(K)} & \colhead{(nepers)} & \colhead{(K \kms)} & \colhead{(\kms)} & \colhead{(\kms)} & \colhead{(K)} & \colhead{(nepers)}\\}
\startdata
 IRAS03282 &  6.37 $\pm$ 0.04 &  6.94 $\pm$ 0.003 &  0.40 $\pm$ 0.006 &  6.0 $\pm$ 0.6 &  4.8 $\pm$  0.5 &  1.40 $\pm$ 0.04 &  7.01 $\pm$ 0.01 &  0.35 $\pm$ 0.02 &  4.2 $\pm$ 0.6 &  2.7 $\pm$ 0.6\\
     HH211 &  5.90 $\pm$ 0.05 &  9.10 $\pm$ 0.003 &  0.45 $\pm$ 0.008 &  4.8 $\pm$ 0.5 &  6.9 $\pm$  0.6 &  2.27 $\pm$ 0.05 &  9.18 $\pm$ 0.01 &  0.38 $\pm$ 0.03 &  4.3 $\pm$ 0.6 &  3.9 $\pm$ 0.8\\
    L1521F &  5.30 $\pm$ 0.02 &  6.47 $\pm$ 0.001 &  0.30 $\pm$ 0.001 &  4.5 $\pm$ 0.1 & 16.9 $\pm$  0.2 &  0.93 $\pm$ 0.03 &  6.56 $\pm$ 0.01 &  0.27 $\pm$ 0.02 &  4.1 $\pm$ 0.2 &  2.2 $\pm$ 0.3\\
 IRAS04325 &  4.31 $\pm$ 0.05 &  5.70 $\pm$ 0.004 &  0.39 $\pm$ 0.011 &  5.3 $\pm$ 1.1 &  4.1 $\pm$  0.8 &  0.03 $\pm$ 0.02 &  5.66 $\pm$ 0.07 &  0.40 $\pm$ 0.11 &  4.7 $\pm$ 1.7 &  0.1 $\pm$ 0.0\\
 IRAS04325N &  6.40 $\pm$ 0.03 &  5.87 $\pm$ 0.003 &  0.33 $\pm$ 0.008 &  6.7 $\pm$ 1.1 &  4.9 $\pm$  0.8 &  0.81 $\pm$ 0.03 &  5.94 $\pm$ 0.01 &  0.29 $\pm$ 0.02 & 21.7 $\pm$ 3.6 &  0.1 $\pm$ 0.0\\
     L1527 &  3.24 $\pm$ 0.03 &  5.90 $\pm$ 0.004 &  0.33 $\pm$ 0.009 &  4.5 $\pm$ 0.8 &  5.9 $\pm$  0.9 & $<$ 0.01 & ... &  ... &  ... & ...\\
     L1527N &  2.87 $\pm$ 0.03 &  5.82 $\pm$ 0.004 &  0.32 $\pm$ 0.009 &  4.2 $\pm$ 0.7 &  6.6 $\pm$  1.0 &  0.22 $\pm$ 0.01 &  5.81 $\pm$ 0.03 &  0.39 $\pm$ 0.08 &  6.4 $\pm$ 1.5 &  0.1 $\pm$ 0.0\\
     RNO43 &  2.50 $\pm$ 0.03 & 10.25 $\pm$ 0.007 &  0.82 $\pm$ 0.019 &  5.5 $\pm$ 2.6 &  0.9 $\pm$  0.4 &  0.30 $\pm$ 0.02 & 10.40 $\pm$ 0.04 &  0.69 $\pm$ 0.08 &  6.2 $\pm$ 3.0 &  0.1 $\pm$ 0.0\\
      L483 & 12.03 $\pm$ 0.06 &  5.35 $\pm$ 0.006 &  0.42 $\pm$ 0.011 &  5.7 $\pm$ 0.2 & 13.7 $\pm$  0.4 &  0.99 $\pm$ 0.07 &  5.39 $\pm$ 0.01 &  0.27 $\pm$ 0.02 &  3.4 $\pm$ 0.5 &  5.5 $\pm$ 1.2\\
     HH108 IRS &  5.47 $\pm$ 0.06 & 10.80 $\pm$ 0.008 &  0.55 $\pm$ 0.017 &  5.0 $\pm$ 1.1 &  4.4 $\pm$  0.9 &  0.61 $\pm$ 0.04 & 10.78 $\pm$ 0.03 &  0.51 $\pm$ 0.08 & 10.1 $\pm$ 2.5 &  0.1 $\pm$ 0.0\\
     HH108 MMS &  2.12 $\pm$ 0.04 & 10.92 $\pm$ 0.005 &  0.38 $\pm$ 0.013 &  4.4 $\pm$ 1.4 &  3.3 $\pm$  1.0 &  0.54 $\pm$ 0.05 & 10.95 $\pm$ 0.04 &  0.50 $\pm$ 0.10 &  4.7 $\pm$ 2.7 &  0.4 $\pm$ 2.8\\
      L673S &  7.30 $\pm$ 0.05 &  6.85 $\pm$ 0.006 &  0.55 $\pm$ 0.015 &  7.6 $\pm$ 2.1 &  2.6 $\pm$  0.7 &  0.74 $\pm$ 0.03 &  6.84 $\pm$ 0.02 &  0.58 $\pm$ 0.06 & 10.6 $\pm$ 3.0 &  0.1 $\pm$ 0.0\\
      L673M &  6.63 $\pm$ 0.05 &  6.90 $\pm$ 0.008 &  0.68 $\pm$ 0.020 &  6.1 $\pm$ 1.4 &  2.9 $\pm$  0.7 &  0.47 $\pm$ 0.04 &  6.93 $\pm$ 0.04 &  0.62 $\pm$ 0.08 &  7.7 $\pm$ 2.0 &  0.1 $\pm$ 0.0\\
      L673N &  5.37 $\pm$ 0.05 &  6.91 $\pm$ 0.009 &  0.70 $\pm$ 0.022 &  5.1 $\pm$ 1.1 &  3.4 $\pm$  0.7 &  0.33 $\pm$ 0.03 &  6.90 $\pm$ 0.04 &  0.47 $\pm$ 0.12 &  3.8 $\pm$ 2.2 &  0.4 $\pm$ 3.3\\
     L1152 &  4.17  $\pm$ 0.04 &  2.64 $\pm$ 0.005 &  0.40 $\pm$ 0.012 &  4.6  $\pm$ 0.9 &  6.0  $\pm$  1.0 &  0.31  $\pm$ 0.03 &  2.71  $\pm$ 0.03 &  0.40  $\pm$ 0.07 &  3.0  $\pm$ 1.1 &  2.6  $\pm$ 1.8\\
     L1157 &  4.85  $\pm$ 0.02 &  2.65 $\pm$ 0.007 &  0.57 $\pm$ 0.017 &  4.6  $\pm$ 0.9 &  4.5  $\pm$  0.8 &  0.78  $\pm$ 0.05 &  2.73  $\pm$ 0.01 &  0.40  $\pm$ 0.02 &  3.6  $\pm$ 0.7 &  2.2  $\pm$ 0.5\\
     L1165 &  1.50  $\pm$ 0.04 & -1.61 $\pm$ 0.006 &  0.43 $\pm$ 0.015 &  3.9  $\pm$ 1.4 &  2.8  $\pm$  1.0 &  0.04  $\pm$ 0.01 & -1.61  $\pm$ 0.04 &  0.26  $\pm$ 0.09 &  5.7  $\pm$ 2.7 &  0.1 $\pm$ 0.0\\

\enddata
\tablecomments{The \nthp\ and \ntdp\ integrated intensities are on the $T_{mb}$ scale and do not include the $\sim$10\% calibration error.
Note that for \ntdp\ $T_{ex}$ is not well constrained when $\tau$ $<$ 0.1 and the uncertainties for $\tau$ = 0.1 are not meaningful.}
\end{deluxetable}

\begin{deluxetable}{llllc}
\tabletypesize{\scriptsize}
\tablewidth{0pt}
\tablecaption{Column densities of \nthp and \ntdp\ Ratios and Offsets\label{param}}
\tablehead{   \colhead{Source} & \colhead{N(\nthp)} & \colhead{N(\ntdp)} & \colhead{N(\ntdp)/N(\nthp)} & \colhead{N2H+/N2D+ offset} \\
   & \colhead{(cm$^{-2}$)} & \colhead{(cm$^{-2}$)} & & \colhead{($^{\prime\prime}$)} }
\startdata
 IRAS03282 &   2.66 $\times$ 10$^{13}$ $\pm$ 3.25 $\times$ 10$^{12}$  &   3.42 $\times$ 10$^{12}$ $\pm$  6.52 $\times$ 10$^{11}$  & 0.129 $\pm$ 0.029 & 8\\
     HH211 &   3.41 $\times$ 10$^{13}$ $\pm$  4.17 $\times$ 10$^{12}$  &   7.57 $\times$ 10$^{12}$ $\pm$  1.58 $\times$ 10$^{12}$  & 0.222 $\pm$ 0.054 & 20\tablenotemark{\dagger}\\
    L1521F &   7.42 $\times$ 10$^{13}$ $\pm$  1.72 $\times$ 10$^{12}$  &   2.09 $\times$ 10$^{12}$ $\pm$  2.00 $\times$ 10$^{11}$  & 0.028 $\pm$ 0.003 & 0\\
 IRAS04325 &   1.51 $\times$ 10$^{13}$ $\pm$  3.97 $\times$ 10$^{12}$  &   $<$ 1.6  $\times$ 10$^{10}$ & $<$ 0.001  &  ... \\
 IRAS04325N &   2.83 $\times$ 10$^{13}$ $\pm$  5.42 $\times$ 10$^{12}$  &   4.05 $\times$ 10$^{11}$ $\pm$  4.89 $\times$ 10$^{10}$  & 0.014 $\pm$ 0.003 & 0  \\
     L1527 &   1.59  $\times$ 10$^{13}$ $\pm$  3.53  $\times$ 10$^{12}$  &   $<$ 1.6  $\times$ 10$^{10}$ & $<$ 0.001 & 35\\
     L1527N &   1.56 $\times$ 10$^{13}$ $\pm$  3.78 $\times$ 10$^{12}$  &   1.85 $\times$ 10$^{11}$ $\pm$  3.59 $\times$ 10$^{10}$  & 0.012 $\pm$ 0.004 & ...\\
     RNO43 &   3.18 $\times$ 10$^{12}$ $\pm$  1.35 $\times$ 10$^{12}$  &   1.79 $\times$ 10$^{11}$ $\pm$  7.02 $\times$ 10$^{10}$  & 0.056 $\pm$ 0.033 & 16\\
      L483 &   1.40 $\times$ 10$^{14}$ $\pm$  6.17 $\times$ 10$^{12}$  &   6.92 $\times$ 10$^{12}$ $\pm$  1.60 $\times$ 10$^{12}$  & 0.049 $\pm$ 0.012 & 16\\
     HH108 IRS & 2.01 $\times$ 10$^{13}$ $\pm$  5.41 $\times$ 10$^{12}$  &   4.03 $\times$ 10$^{11}$ $\pm$  8.27 $\times$ 10$^{10}$  & 0.020 $\pm$ 0.007 & 0\tablenotemark{a}\\
     HH108 MMS & 5.99 $\times$ 10$^{12}$ $\pm$  2.55 $\times$ 10$^{12}$  &   5.32 $\times$ 10$^{11}$ $\pm$  7.23 $\times$ 10$^{11}$  & 0.089 $\pm$ 0.127 & ...\\
      L673S &   1.89 $\times$ 10$^{13}$ $\pm$  5.05 $\times$ 10$^{12}$  &   3.47 $\times$ 10$^{11}$ $\pm$  5.89 $\times$ 10$^{10}$  & 0.018 $\pm$ 0.006 & 18\\
      L673M &   1.77 $\times$ 10$^{13}$ $\pm$  4.56 $\times$ 10$^{12}$  &   2.54 $\times$ 10$^{11}$ $\pm$  4.84 $\times$ 10$^{10}$  & 0.014 $\pm$ 0.005 & ...\\
      L673N &   1.57 $\times$ 10$^{13}$ $\pm$  4.05 $\times$ 10$^{12}$  &   2.60 $\times$ 10$^{11}$ $\pm$  3.98 $\times$ 10$^{11}$  & 0.017 $\pm$ 0.026 & ... \\
    L1152 &   2.09 $\times$ 10$^{13}$ $\pm$  5.20 $\times$ 10$^{12}$  &   1.49 $\times$ 10$^{12}$ $\pm$  8.79 $\times$ 10$^{11}$  & 0.071 $\pm$ 0.046 & 0\\
     L1157 &   1.82 $\times$ 10$^{13}$ $\pm$  4.49 $\times$ 10$^{12}$  &   2.19 $\times$ 10$^{12}$ $\pm$  5.58 $\times$ 10$^{11}$  & 0.120 $\pm$ 0.043 & 3.5\\
     L1165 &   3.76 $\times$ 10$^{12}$ $\pm$  1.97 $\times$ 10$^{12}$  &   3.75 $\times$ 10$^{10}$ $\pm$  2.03 $\times$ 10$^{10}$  & 0.010 $\pm$ 0.008& 8\\

\enddata
\tablecaption{\nthp\ and \ntdp\ column densities and ratios are taken toward the position of the 
protostar and are the average of all spectra in a 24\arcsec\ radius centered on the protostar position.
L1527N is taken at (0\arcsec,60\arcsec), and L673N,M, and S refer to north, middle, and south protostars
respectively.}
\tablenotetext{\dagger}{Taken from the column density maps, emission maps are spatially coincident.}
\tablenotetext{a}{Shift between protostar and \nthp\ is observed even though \nthp\ and \ntdp\ are spatially coincident.}
\end{deluxetable}

\end{document}